\documentclass{jfm}
\usepackage{graphicx}
\usepackage{epstopdf, epsfig}

\usepackage{amsmath}
\usepackage{newtxmath}
\usepackage{hyperref}
\usepackage{makecell}

\shorttitle{Taylor dispersion in arbitrarily shaped axisymmetric channels}
\shortauthor{R. Chang and J. G. Santiago}

\title{
Taylor dispersion in arbitrarily shaped axisymmetric channels
}

\author{Ray Chang\aff{1}
 \and Juan G. Santiago\aff{2} \corresp{\email{juan.santiago@stanford.edu}}
 }
\affiliation{\aff{1}Department of Bioengineering, Stanford University, Stanford, CA, 94305, United States
\aff{2}Department of Mechanical Engineering, Stanford University, Stanford, CA, 94305, United States}

\begin{document}
\maketitle

\begin{abstract}
Advective dispersion of solutes in long thin axisymmetric channels is important to the analysis and design of a wide range of devices, including chemical separation systems and microfluidic chips. Despite extensive analysis of Taylor dispersion in various scenarios, most studies focused on long-term dispersion behavior and cannot capture the transient evolution of solute zone across the spatial variations in the channel. In the current study, we analyze the Taylor-Aris dispersion for arbitrarily shaped axisymmetric channels. We derive an expression for solute dynamics in terms of two coupled ordinary differential equations (ODEs). These two ODEs allow prediction of the time evolution of the mean location and axial (standard deviation) width of the solute zone as a function of the channel geometry. We compare and benchmark our predictions with Brownian dynamics simulations for a variety of cases including linearly expanding/converging channels and periodic channels. We also present an analytical description of the physical regimes of transient positive versus negative axial growth of solute width. Finally, to further demonstrate the utility of the analysis, we demonstrate a method to engineer channel geometries to achieve desired solute width distributions over space and time. We apply the latter analysis to generate a geometry that results in a constant axial width and a second geometry that results in a sinusoidal axial variance in space.
\end{abstract}


\section{Introduction}
Advective dispersion of solutes in long thin tubes is important to the design and optimization of a wide range of devices, from chemical processing and separations systems to microfluidic chips \citep{Brenner1993}. G. I. Taylor first reported a closed-form solution for the long-term dispersive behavior of a solute in a circular cylindrical tube driven by a fully-developed Poiseuille flow \citep{Taylor1953}. Taylor's original solution considered a regime where axial molecular diffusion (molecular diffusion in the axial direction) is negligible compared to the dispersion effect, but radial molecular diffusion continues to play a role because of the radial concentration gradient from the convection. In terms of inequalities, this original analysis is applicable for $L/a\gg Pe_a\gg6.9$, where $L$ is the characteristic channel length, $a$ is the radius of the tube, and $Pe_a$ is a Peclet number based on $a$. Aris later included the effects of axial molecular diffusion of the solute and introduced the method of moments which enabled treatment of all stages of the dispersive process in long-thin channels with fairly arbitrary cross-sections \citep{Aris1956}. Dispersion analyses including molecular diffusion and dynamic regimes where radial diffusion times are much shorter than advection times are typically termed Taylor-Aris analyses. Since these seminal papers, there has been much work analyzing the dispersion behavior in various systems. The book of Brenner \& Edwards includes a broad range example analyses including dispersion of flows through porous media and effects of chemical reactions and surface adsorption \citep{Brenner1993}. 

A significant emphasis has been placed on the dispersion problem in spatially-periodic channel geometries. The seminal work of H. Brenner (which is later referred as Brenner-Aris theory, or the macrotransport paradigm) \citep{Brenner1980} provided a generalized framework to predict the long-term dispersion behavior of point-size particles in any spatially periodic channels or porous media. This approach solves an elliptical partial differential equation of a cell field \textbf{B} (commonly referred as $\textbf{B}$ field) defined on a periodic unit cell and uses this to compute a long-term dispersion tensor. Hoagland \& Prud'homme \citep{Hoagland1985} applied Brenner-Aris theory to the dispersion in a long-thin, axisymmetric channel with a sinusoidal variation of radius along the streamwise direction. They presented numerical solutions of the method of moments for arbitrary wavelengths, and they also derived an analytical expression for the long-term dispersion of solute in the limit of long wavelength which compared well with their numerical model. Bolster et al. performed a similar analysis to Hoagland \& Prud'Homme in axisymmetric channel with a sinusoidal variation of radius, and extended the results by comparing the predictions with Brownian dynamics simulations \citep{Bolster2009}. Dorfman and his coworkers also used Brenner's method but for the advective component considered electrophoretic and electro-osmotic flows with finite double layers \citep{Yariv2007, Dorfman2008, Dorfman2009}. Adrover et al. later considered dispersion of both point and finite-sized particles in a long-thin sinusoidal channel \citep{Adrover2019}. For point particles, they obtained various asymptotic expressions for the effective, long-term dispersion coefficient in the limits of large and small $Pe_a$.

Methods other than Brenner's theory were also proposed to analyze the dispersion in periodic channels. For example, Rosencrans \citep{Rosencrans1997} followed the center manifold theory of Mercer and Roberts \citep{Mercer1990} and derived the dispersion coefficient for periodic curved channels. They hypothesized that their analysis extended to strong variations in geometry but did not support this with numerical analyses. In a series of papers, Martens and his coworkers explored the dispersion process in periodic channels from the (Lagrangian) perspectives of individual particles and formulated dispersion theory based on the Fick-Jacobs equation \citep{Martens2011,Martens2013, Martens2013b, Martens2014}. 

Several studies also addressed the dispersion problem in spatially non-periodic systems. Early work from Saffman analyzed the long-term longitudinal dispersion in the direction of mean flow in a random porous media where the flow satisfies Darcy's law \citep{Saffman1959}. Gill, Ananthakrishnan and Nunge analytically and numerically analyzed the dispersion in a velocity entrance region where the flow field is not fully developed \citep{Gill1968}. Gill and Guceri later investigated numerically the dispersion in diverging channels over a wide range of angles of divergence and Peclet number \citep{Gill1971}. Smith described the longitudinal dispersion in a varying channel in terms of the memory effect of the dispersion coefficient. Smith also investigated the changes in dispersion coefficients in several profiles that are pertinent to geophysics, such as sudden changes in breadth, centrifugally driven secondary flow associated with the curvature in the flow path, and changes in the depth profile \citep{Smith1983}. Mercer and Roberts used center manifold theory to derive the spatially dependent dispersion coefficient in channels with varying flow properties. Although they claimed that their approach may be applicable to time-dependent flow and variable diffusivity, the dispersion coefficient alone can be misleading in predicting actual dispersion behavior \citep{Mercer1990}. Horner, Arce and Locke studied the Taylor dispersion in an axisymmetric system with two linear profiles of radius changes, which was intended to approximate the convergent and divergent sections of stenosis sites in arteries  \citep{Horner1995}. Bryden \& Brenner analyzed the Taylor-Aris dispersion in a diverging conical channel and a flared, axisymmetric Venturi tube geometry \citep{Bryden1996}. They used multiple-timescale analysis to obtain an asymptotic expression for the effective dispersion coefficient and derived a partial differential equation (PDE) for a conditional probability density function describing the solute. However, the latter work neither provided a solution for this PDE nor analyzed the long-term dispersion and growth of a solute zone. Stone and Brenner also investigated the dispersion in a different spatially non-periodic system. The latter work considered dispersion in a radially expanding flow between large parallel plates. The latter flow results in a mean velocity which varies in the streamwise direction and therefore exhibits a location-dependent dispersion coefficient \citep{Stone1999}.

Note that all the aforementioned literature considers the spatial dispersion problem, as the dispersion process is quantified by the spatial moments of the averaged solute distribution. An alternative framework of temporal dispersion problem, proposed by Dankwerts \citep{Danckwerts1953}, quantifies the dispersion process in terms of its temporal moments. This formulation is particularly useful when the solute is monitored at a given distance from the inlet, and can also be useful in analyzing and benchmarking results from numerical models \citep{Rodrigues2021}. Although the formulations are different, the spatial moments and temporal moments of the dispersion process have been shown to be interrelated \citep{Vikhansky2014, Ginzburg2018}.

To our knowledge, all previous analyses focused on the long-term dispersion behavior and were not able to provide a simple prediction of both the short-term and long-term spatial evolution of solute. We also know of no analytical work toward Taylor-Aris dispersion in an arbitrarily-shaped axisymmetric channel. Such analyses have significant potential to influence the design and analyses of a wide range of systems, including microfluidic devices. In the current study, we analyze the Taylor-Aris dispersion in an arbitrarily-shaped axisymmetric channel with a slowly varying radius. We derive an expression for solute dynamics in terms of two coupled ordinary differential equations (ODEs). These two ODEs enable prediction of the time evolution of the solute zone based on channel geometry and the assumed lubrication flow. We compare and benchmark our predictions with Brownian dynamics analysis. We further develop a formulation useful in inverse problems where channels are designed to achieve desired spatial distribution of solute variance. We demonstrate this for the design of two channel geometries of specialized function. The first geometry results in a constant axial variance and the second results in a sinusoidal axial variance in space.

\section{Theory}
\subsection{Taylor-Aris dispersion in arbitrarily shaped cylinders}

\begin{table}
\small
\caption{Summary of assumptions made in this study.} 
\centering 
\begin{tabular}{ c } 
\\[0cm]
\makecell[l]{\textbf{Key parameters used for scaling analysis:}\\$a_0$: radius of the channel at $x=0$, taken as the characteristic radius.\\$U_0$: cross-sectional averaged axial velocity at $x=0$, taken as the characteristic axial velocity.\\$\lambda$: characteristic wavelength of spatial variation of the channel.\\\hspace{0.3cm}$\mathcal{O}(\beta)\sim\mathcal{O}(a_0/\lambda)$. $\mathcal{O}(\gamma)\sim\mathcal{O}(a_0/\lambda^2)$.\\$\sigma$: characteristic axial dimension of the solute zone.\\$c_0$: characteristic magnitude of the cross-sectional average concentration of solute.\\$c_0^{\prime}$: characteristic magnitude of deviation concentration of solute.\\$Re$: characteristic Reynolds number.\\$Pe_a$: local Peclet number defined as $Ua/D$. Its quantity at $x=0$ is $Pe_{a_0}$.}\\
\\[0cm]
\makecell[l]{\textbf{4 smallness parameters:}\\(1): $\epsilon=a_0/\lambda\ll1$: slowly varying channel, or long-wavelength assumptions.\\(2): $\eta=a_0/\sigma\ll1$.\\(3): $c_0^{\prime}/c_0\ll1$: small deviation in concentration from cross-sectional average.\\(4): $\sigma/\lambda\ll1$: solute zone does not spread across multiple spatial variation.} \\ 
\\[0cm]
\makecell[l]{\textbf{Additional assumptions and asymptotics needed to obtain Eq. \ref{eq:taylor_final}:}\\(1) lubrication flow: $\epsilon^2Re\ll1$. \\(2): $t_{\mathrm{obs}}\sim\sigma/U_0$.\\(3): Taylor ansatz: On the left-hand side (LHS) of Eq. \ref{eq:pert}, only keep terms on order\\\hspace{0.3cm} $\mathcal{O}(U_0c_0/\sigma)$.\\(4): Taylor ansatz: On the right-hand side (RHS) of Eq. \ref{eq:pert}: only keep terms on order\\\hspace{0.3cm} $\mathcal{O}(Dc_0^{\prime}/a_0^2)$.\\(5): In Equation \ref{eq:evaluated_c_prime}, only keep terms on the order of $\frac{U_0c_0}{\sigma}\mathcal{O}(\eta Pe_{a_{0}})$ and $\frac{Dc_0}{\sigma^2}\mathcal{O}(\frac{\sigma}{\lambda})$.\\\hspace{0.3cm}The associated error magnitude is $\frac{U_0c_0}{\sigma}\mathcal{O}(\eta\epsilon)$ and $\frac{Dc_0}{\sigma^2}\mathcal{O}(\epsilon^2)$.}\\
\\[0cm]
\makecell[l]{\textbf{Additional assumptions and asymptotics needed to derive Eq. \ref{eq:x_bar} and \ref{eq:var}}\\\textbf{from Eq. \ref{eq:taylor_final}:}\\(1): Taylor expansion: functions $a^{-2}$, $\beta/a$, $x/a^2$, $(x-\overline{x})\beta/a$ can be described by Taylor series \\\hspace{0.3cm}at $x=\overline{x}$.\\(2): For Equation \ref{eq:x_bar}, keep terms on order $\mathcal{O}(U_0, D/\lambda)$. The associated error magnitude is\\\hspace{0.3cm} $\mathcal{O}(U_0(\frac{\sigma}{\lambda})^2)$ and $\mathcal{O}(\frac{D}{\lambda}(\frac{\sigma}{\lambda})^2)$.\\(3): For Equation \ref{eq:var}, keep terms on order $D\mathcal{O}(1, Pe_{a_0}^2, Pe_{a_0}\frac{\sigma^2}{a_0\lambda})$. The associated error\\\hspace{0.3cm} magnitude is $D\mathcal{O}((\frac{\sigma}{\lambda})^2, Pe_{a_0}^2(\frac{\sigma}{\lambda})^2, Pe_{a_0}\frac{\sigma^2}{a_0\lambda}\frac{\sigma}{\lambda}\mathrm{Skew}_x)$}\\
\\[0cm]
\makecell[l]{\textbf{Assumptions needed to derive Eq. \ref{eq:arb_var} from Eq. \ref{eq:var}:}\\(1): $\frac{\mathrm{d}}{\mathrm{d}t}\simeq\frac{\mathrm{d}\overline{x}}{\mathrm{d}t}\frac{\mathrm{d}}{\mathrm{d}x}$}\\
\\[0cm]
\end{tabular}
\label{table:assumptions_summary} 
\end{table}

\begin{figure}
\centerline{\includegraphics[width=1.0\textwidth]{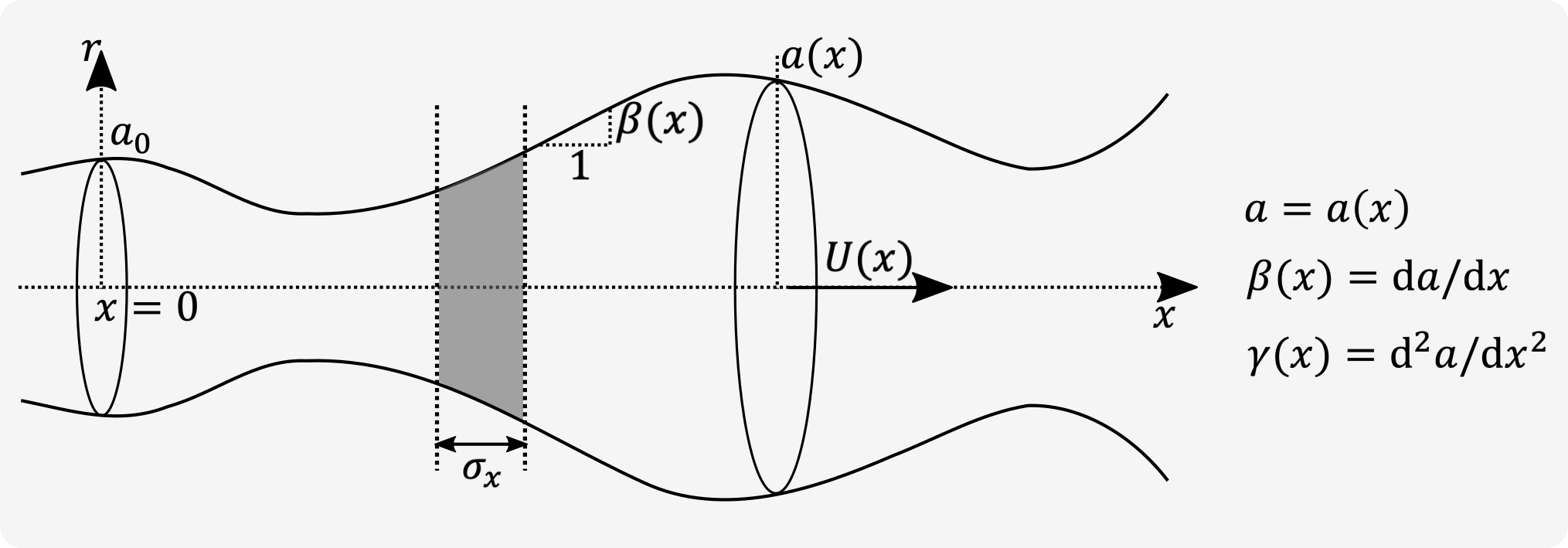}}
\caption{Schematic of an axisymmetric channel with a slowly varying, arbitrary distribution of radius, $a(x)$. A nominal radius is taken as $a_0$ at $x=0$. The slope and the curvature of the cylinder wall are respectively $\beta(x)$ and $\gamma(x)$, as shown. $\sigma_{x}$ is the characteristic width of a solute zone and $U(x)$ is the area-averaged axial velocity distribution. The ($x$-direction) length of the channel depicted schematically in the sketch has been compressed relative to its characteristic radius for clarity of presentation.
}\label{fig:scheme}
\end{figure}

We analyze Taylor-Aris dispersion for flow in an axisymmetric channel with a slowly varying, arbitrary radius $a = a(x)$. We define variables for the first and second derivatives as $\beta(x)=\text{d}a/\text{d}x$ and $\gamma(x)=\text{d}^2a/\text{d}x^2$. Given the azimuthal symmetry, the concentration $c(x,r,t)$ of a solute evolves according to the following convection-diffusion equation in cylindrical coordinates.
\begin{equation}
\frac{\partial c}{\partial t} + u_{r}\frac{\partial c}{\partial r} + u_{x}\frac{\partial c}{\partial x} = D\left(\frac{1}{r}\frac{\partial}{\partial r}\left(r\frac{\partial c}{\partial r}\right) + \frac{\partial^2 c}{\partial x^2}\right)
\end{equation}
For the velocity field, the Navier-Stokes equation is solved with the assumptions typical of lubrication theory. That is, we consider a slowly varying radius with a characteristic radius much smaller than the axial distance of interest \citep{langlois1964slow}. Hence we write the velocity field as
\begin{equation}\label{eq:lubrication}
\begin{split}
u_{x}(x,r)=2U(x)\left(1-\frac{r^2}{a(x)^2}\right)+\mathcal{O}(U_0Re\epsilon^2, U_0\epsilon^2)\\
u_r(x,r)=2\beta(x)U(x)\left(\frac{r}{a(x)}-\frac{r^3}{a(x)^3}\right)+\mathcal{O}(U_0Re\epsilon^3, U_0\epsilon^3),
\end{split}
\end{equation}
where $Re$ is the Reynolds number, $\epsilon$ is our primary smallness parameter defined as the ratio between the characteristics radius $a_0$ (defined as the radius at $x=0$) and the characteristic axial length scale of radius variation $\lambda$. The second terms on the right-hand side (RHS) indicate the order of magnitude of the truncation error associated with this asymptotic approximation. $U_0$ is the characteristic mean axial velocity defined here as the mean axial velocity at $x=0$. We also assume that the characteristic width of the solute zone $\sigma$ is much smaller than the characteristic wavelength of radius variation, or $\sigma/\lambda\ll1$. For the Taylor-Aris analysis, we follow the notation of Stone and Brenner \citep{Stone1999} and expand each variable into cross-sectional averages of the form $\langle(\cdot)\rangle\equiv\frac{1}{\pi a(x)^2}\int_{0}^{a(x)}2\pi r(\cdot)\text{d}r$ and deviations therefrom defined as $(\cdot)^{\prime} \equiv (\cdot) - \langle(\cdot)\rangle$. Whence, the area-averaged velocity components are
\begin{equation}
\langle u_{x}\rangle=U(x),\quad \langle u_r\rangle=\frac{8}{15}\beta(x)U(x)
\end{equation}
\begin{equation}
u_x^{\prime}=U(x)\left(1-\frac{2r^2}{a(x)^2}\right),\quad u_r^{\prime}=2\beta(x)U(x)\left(\frac{r}{a(x)}-\frac{r^3}{a(x)^3}-\frac{4}{15}\right)
\end{equation}
We then expand the convective-diffusion equation as
\begin{equation}\label{eq:full}
\begin{split}
\frac{\partial \langle c\rangle}{\partial t}+\frac{\partial c^{\prime}}{\partial t} + \langle u_{r}\rangle\frac{\partial c^{\prime}}{\partial r} + u_{r}^{\prime}\frac{\partial c^{\prime}}{\partial r} + U(x)\frac{\partial \langle c\rangle}{\partial x} + U(x)\frac{\partial c^{\prime}}{\partial x} + u_{x}^{\prime}\frac{\partial \langle c\rangle}{\partial x} + u_{x}^{\prime}\frac{\partial  c^{\prime}}{\partial x} \\= D\left(\frac{1}{r}\frac{\partial}{\partial r}\left(r\frac{\partial c^{\prime}}{\partial r}\right) + \frac{\partial^2 \langle c\rangle}{\partial x^2} + \frac{\partial^2 c^{\prime}}{\partial x^2}\right)
\end{split}
\end{equation}
The area-average of the latter equation is then
\begin{equation}\label{eq:ave}
\begin{split}
\frac{\partial \langle c\rangle}{\partial t} + \left\langle u_{r}\right\rangle\left\langle\frac{\partial c^{\prime}}{\partial r}\right\rangle + \left\langle u_{r}^{\prime}\frac{\partial c^{\prime}}{\partial r}\right\rangle + U(x)\frac{\partial \langle c\rangle}{\partial x} + U(x)\left\langle\frac{\partial c^{\prime}}{\partial x}\right\rangle + \left\langle u_{x}^{\prime}\frac{\partial  c^{\prime}}{\partial x}\right\rangle \\= D\left(\left\langle\frac{1}{r}\frac{\partial}{\partial r}\left(r\frac{\partial c^{\prime}}{\partial r}\right)\right\rangle + \frac{\partial^2 \langle c\rangle}{\partial x^2} + \left\langle\frac{\partial^2 c^{\prime}}{\partial x^2}\right\rangle\right)
\end{split}
\end{equation}
We simplify four of these terms in Eq. \ref{eq:ave} using Leibniz's rule, integration by parts, and chain rule as follows:
\begin{equation}
\begin{split}
\left\langle\frac{\partial c^{\prime}}{\partial r}\right\rangle &= \frac{1}{a(x)^2}\int_{0}^{a(x)}2r\frac{\partial c^{\prime}}{\partial r}\mathrm{d}r=\frac{2}{a(x)^2}\int_{0}^{a(x)}\left(\frac{\partial}{\partial r}(rc^{\prime})-c^{\prime}\right)\mathrm{d}r\\&=\frac{2}{a(x)^2}\left(a\left.c^{\prime}\right\vert_{r=a}-\int_{0}^{a}c^{\prime}\mathrm{d}r\right)\\
\end{split}
\end{equation}
\begin{equation}
\begin{split}
\left\langle\frac{\partial c^{\prime}}{\partial x}\right\rangle &= \frac{1}{a(x)^2}\int_{0}^{a(x)}2r\frac{\partial c^{\prime}}{\partial x}\mathrm{d}r = \frac{2}{a(x)^2}\int_{0}^{a(x)}\frac{\partial (rc^{\prime})}{\partial x}\mathrm{d}r\\&=\frac{2}{a(x)^2}\left[\frac{\mathrm{d}}{\mathrm{d}x}\int_{0}^{a(x)}rc^{\prime}\mathrm{d}r-a\left.c^{\prime}\right\vert_{r=a}\beta(x)\right]=\frac{-2\left.c^{\prime}\right\vert_{r=a}\beta(x)}{a(x)}\\
\end{split}
\end{equation}
\begin{equation}
\begin{split}
\left\langle\frac{1}{r}\frac{\partial}{\partial r}\left(r\frac{\partial c^{\prime}}{\partial r}\right)\right\rangle &= \frac{2}{a(x)^2}\int_{0}^{a(x)}r\frac{1}{r}\frac{\partial}{\partial r}\left(r\frac{\partial c^{\prime}}{\partial r}\right)\mathrm{d}r=\frac{2}{a(x)^2}\left(a\left.\frac{\partial c^{\prime}}{\partial r}\right\vert_{r=a} - 0\right)\\ &= \frac{2}{a}\left.\frac{\partial c^{\prime}}{\partial r}\right\vert_{r=a}\\
\end{split}
\end{equation}
\begin{equation}
\begin{split}
\left\langle\frac{\partial^2 c^{\prime}}{\partial x^2}\right\rangle &= \frac{2}{a(x)^2}\int_{0}^{a(x)}r\frac{\partial^2c^{\prime}}{\partial x^2}\mathrm{d}r=\frac{2}{a(x)^2}\int_{0}^{a(x)}\frac{\partial^2 (rc^{\prime})}{\partial x^2}\mathrm{d}r\\&=\frac{2}{a^2}\left[\frac{\mathrm{d}}{\mathrm{d}x}\int_{0}^{a(x)}\frac{\partial (rc^{\prime})}{\partial x}\mathrm{d}r - \beta(x)\frac{\partial(a\left.c^{\prime}\right\vert_{r=a})}{\partial x}\right]\\&=\frac{2}{a^2}\left[\frac{\mathrm{d}}{\mathrm{d}x}\left(\frac{\mathrm{d}}{\mathrm{d}x}\int_{0}^{a(x)}rc^{\prime}\mathrm{d}r -\beta(x)a\left.c^{\prime}\right\vert_{r=a}\right) - \beta(x)\frac{\partial(a\left.c^{\prime}\right\vert_{r=a})}{\partial x}\right]\\&=-\frac{2\gamma \left.c^{\prime}\right\vert_{r=a}}{a} - \frac{4\beta}{a^2}\frac{\partial}{\partial x}(a\left.c^{\prime}\right\vert_{r=a})
\end{split}
\end{equation}
Inserting these terms to equation \ref{eq:ave}, we have
\begin{equation}\label{eq:ave_plug}
\begin{split}
\frac{\partial \langle c\rangle}{\partial t} + \frac{2\langle u_{r}\rangle}{a^2}\left(a\left.c^{\prime}\right\vert_{r=a}-\int_{0}^{a}c^{\prime}\mathrm{d}r\right) + \left\langle u_{r}^{\prime}\frac{\partial c^{\prime}}{\partial r}\right\rangle + U(x)\frac{\partial \langle c\rangle}{\partial x} -\frac{2U(x)\left.c^{\prime}\right\vert_{r=a}\beta(x)}{a(x)} + \left\langle u_{x}^{\prime}\frac{\partial  c^{\prime}}{\partial x}\right\rangle \\= \frac{2D}{a}\left.\frac{\partial c^{\prime}}{\partial r}\right\vert_{r=a} + D\frac{\partial^2 \langle c\rangle}{\partial x^2} - \frac{4\beta D}{a^2}\frac{\partial}{\partial x}(a\left.c^{\prime}\right\vert_{r=a})-\frac{2\gamma D\left.c^{\prime}\right\vert_{r=a}}{a}
\end{split}
\end{equation}
Subtracting Equation \ref{eq:ave_plug} from Equation \ref{eq:full} yields the following equation for the deviation of concentration:
\begin{equation}\label{eq:pert}
\begin{split}
\frac{\partial c^{\prime}}{\partial t} + \langle u_{r}\rangle\frac{\partial c^{\prime}}{\partial r} - \frac{2\langle u_{r}\rangle}{a^2}\left(a\left.c^{\prime}\right\vert_{r=a}-\int_{0}^{a}c^{\prime}\mathrm{d}r\right) + u_{r}^{\prime}\frac{\partial c^{\prime}}{\partial r} -  \left\langle u_{r}^{\prime}\frac{\partial c^{\prime}}{\partial r}\right\rangle\\
+ U(x)\frac{\partial c^{\prime}}{\partial x} + \frac{2U(x)\left.c^{\prime}\right\vert_{r=a}\beta(x)}{a(x)} + u_{x}^{\prime}\frac{\partial \langle c\rangle}{\partial x} + u_{x}^{\prime}\frac{\partial  c^{\prime}}{\partial x} - \left\langle u_{x}^{\prime}\frac{\partial  c^{\prime}}{\partial x}\right\rangle \\
= D\frac{1}{r}\frac{\partial}{\partial r}\left(r\frac{\partial c^{\prime}}{\partial r}\right) -  \frac{2D}{a}\left.\frac{\partial c^{\prime}}{\partial r}\right\vert_{r=a} + D\frac{\partial^2 c^{\prime}}{\partial x^2} + \frac{4\beta D}{a^2}\frac{\partial}{\partial x}(a\left.c^{\prime}\right\vert_{r=a})+\frac{2\gamma D\left.c^{\prime}\right\vert_{r=a}}{a}
\end{split}
\end{equation}
To determine the dominant terms on the left and right hand side of the equation, we perform a scaling analysis with the following scales: $\langle c\rangle=c_0\langle c^{*}\rangle$, $c^{\prime}=c_0^{\prime}c^{\prime *}$, $x=\sigma x^{*}$, $r=a_0r^{*}$, $\eta=a_0/\sigma$, $t=t_{\mathrm{obs}}t^{*}=\frac{\sigma}{U_0}t^{*}$, $u_x^{\prime}=U_0u_x^{\prime *}$, and $u_r=\epsilon U_0u_r^{*}$. We define $t_{\mathrm{obs}}$ as the characteristic time over which the solute is observed. This observation time scale is analogous to the advective time scale of traditional Taylor-Aris theory. $c_0$ and $c_0^{\prime}$ are the characteristic scales of the area-averaged solute concentration and its deviation, respectively. Our scaling assumes 4 smallness parameters: $\epsilon=a_0/\lambda\ll1$, $\eta=a_0/\sigma\ll1$, $\sigma/\lambda\ll1$, and $c_0^{\prime}/c_0\ll1$. We summarize the order of magnitude of each term in Equation \ref{eq:pert} in Table \ref{table:eq12_summary}.

\begin{table}
\small
\caption{Ranked summary of terms in Equation \ref{eq:pert}.} 
\centering 
\begin{tabular}{ c | c } 
\hline
\multicolumn{2}{l}{\textbf{Left-hand side (LHS)}}\\
\hline
\makecell[l]{$\frac{U_0c_0}{\sigma}\times\mathcal{O}\left(1\right)$} & \makecell[l]{$u_{x}^{\prime}\frac{\partial \langle c\rangle}{\partial x}$}\\
\\[0cm]
\makecell[l]{$\frac{U_0c_0}{\sigma}\times\mathcal{O}\left(\frac{c_0^{\prime}}{c_0}\right)$} & \makecell[l]{$\frac{\partial c^{\prime}}{\partial t}$, $U\frac{\partial c^{\prime}}{\partial x}$, $u_{x}^{\prime}\frac{\partial  c^{\prime}}{\partial x}$, $\left\langle u_{x}^{\prime}\frac{\partial  c^{\prime}}{\partial x}\right\rangle$}\\
\\[0cm]
\makecell[l]{$\frac{U_0c_0}{\sigma}\times\mathcal{O}\left(\frac{c_0^{\prime}}{c_0}\frac{\sigma}{\lambda}\right)$} & \makecell[l]{$\langle u_{r}\rangle\frac{\partial c^{\prime}}{\partial r}$, $\frac{2\langle u_{r}\rangle}{a^2}\left(a\left.c^{\prime}\right\vert_{r=a}-\int_{0}^{a}c^{\prime}\mathrm{d}r\right)$, $u_{r}^{\prime}\frac{\partial c^{\prime}}{\partial r}$, $\left\langle u_{r}^{\prime}\frac{\partial c^{\prime}}{\partial r}\right\rangle$, $\frac{2U\left.c^{\prime}\right\vert_{r=a}\beta(x)}{a(x)}$}\\
\hline
\multicolumn{2}{l}{\textbf{Right-hand side (RHS)}}\\
\hline
\makecell[l]{$\frac{Dc_0^{\prime}}{a_0^2}\times\mathcal{O}\left(1\right)$} & \makecell[l]{$D\frac{1}{r}\frac{\partial}{\partial r}\left(r\frac{\partial c^{\prime}}{\partial r}\right)$, $\frac{2D}{a}\left.\frac{\partial c^{\prime}}{\partial r}\right\vert_{r=a}$}\\
\\[0cm]
\makecell[l]{$\frac{Dc_0^{\prime}}{a_0^2}\times\mathcal{O}\left(\eta^2\right)$} & \makecell[l]{$D\frac{\partial^2 c^{\prime}}{\partial x^2}$}\\
\\[0cm]
\makecell[l]{$\frac{Dc_0^{\prime}}{a_0^2}\times\mathcal{O}\left(\eta\epsilon\right)$} & \makecell[l]{$\frac{4\beta D}{a^2}\frac{\partial}{\partial x}(a\left.c^{\prime}\right\vert_{r=a})$}\\
\\[0cm]
\makecell[l]{$\frac{Dc_0^{\prime}}{a_0^2}\times\mathcal{O}\left(\epsilon^2\right)$} & \makecell[l]{$\frac{2\gamma D\left.c^{\prime}\right\vert_{r=a}}{a}$}\\
\hline
\end{tabular}
\label{table:eq12_summary} 
\end{table}

Consistent with a Taylor-Aris dispersion regime, keeping the dominant advective dispersion term and the dominant radial diffusion term results in an approximate balance as
\begin{equation}\label{eq:balanceRHS}
    u_{x}^{\prime}\frac{\partial \langle c\rangle}{\partial x}=D\frac{1}{r}\frac{\partial}{\partial r}\left(r\frac{\partial c^{\prime}}{\partial r}\right) -  \frac{2D}{a}\left.\frac{\partial c^{\prime}}{\partial r}\right\vert_{r=a}
\end{equation}
Next, we consider the boundary condition on the inner wall of the channel. We impose a no flux boundary condition at each axial location along the wall, $\bigtriangledown c\cdot \hat{n}=0$. This requires $-\left.\frac{\partial c}{\partial r}\right\vert_{r=a}+\beta\left.\frac{\partial c}{\partial x}\right\vert_{r=a}=-\left.\frac{\partial c^{\prime}}{\partial r}\right\vert_{r=a}+\beta\frac{\partial \langle c\rangle}{\partial x}+\beta\left.\frac{\partial c^{\prime}}{\partial x}\right\vert_{r=a}=0$. Substituting the boundary condition into Equation \ref{eq:balanceRHS} and keeping only terms of the same order as the dominant terms in Table \ref{table:eq12_summary}, we have
\begin{equation}\label{eq:balance}
\begin{split}
D\frac{1}{r}\frac{\partial}{\partial r}\left(r\frac{\partial c^{\prime}}{\partial r}\right)\approx u_{x}^{\prime}\frac{\partial \langle c\rangle}{\partial x} + \frac{2D\beta}{a}\frac{\partial \langle c\rangle}{\partial x}.
\end{split}
\end{equation}

We next directly integrate Equation \ref{eq:balance} as
\begin{equation}
r\frac{\partial c^{\prime}}{\partial r} = \frac{\partial \langle c\rangle}{\partial x}\frac{U(x)}{D}\left(\frac{r^2}{2} - \frac{r^4}{2a^2}\right) +\frac{\beta r^2}{a}\frac{\partial \langle c\rangle}{\partial x} + c_1(x,t)
\end{equation}
Here, the function $c_1(x,t)$ results from the partial integration with respect to $r$. Evaluating this expression at $r=0$, we see $r\left.\frac{\partial c^{\prime}}{\partial r}\right\vert_{r=0}=c_1=0$, whence $c_1(x,t)=0$. Integrating the equation a second time we obtain
\begin{equation}
c^{\prime} = \frac{\partial \langle c\rangle}{\partial x}\frac{U(x)}{D}\left(\frac{r^2}{4} - \frac{r^4}{8a^2} \right) +\frac{\beta}{2a}r^2\frac{\partial \langle c\rangle}{\partial x} + c_2(x,t)
\end{equation}
By the definition of the fluctuating quantity, $\int_0^{a}rc^{\prime}\mathrm{d}r=0$. We use this to solve for $c_2$ and obtain an expression of $c^{\prime}$
\begin{equation}\label{eq:c_prime}
c^{\prime}(x,r,t) = \frac{\partial \langle c\rangle}{\partial x}\frac{U(x)}{D}\left(\frac{r^2}{4} - \frac{r^4}{8a^2} -\frac{a^2}{12}\right)+\left(\frac{\beta r^2}{2a} - \frac{1}{4}\beta a\right)\frac{\partial \langle c\rangle}{\partial x},
\end{equation}
with error of order of $c_0\times\mathcal{O}\left(Pe_{a_0}\frac{c_0^{\prime}}{c_0}\eta, Pe_{a_0}Re\epsilon^2\eta, Pe_{a_0}\epsilon^2\eta \right)$.

We next differentiate Eq. \ref{eq:c_prime} with respect to $x$, multiply by the known function $u_x^{\prime}$, and take an area integral to construct the term $\left\langle u_x^{\prime}\frac{\partial c^{\prime}}{\partial x}\right\rangle$. We perform similar procedures for each term in Eq. \ref{eq:ave_plug}, and we arrive at Equation \ref{eq:evaluated_c_prime}. We rank the terms in Eq. \ref{eq:evaluated_c_prime} and summarize in Table \ref{table:eq11_summary}.
\begin{equation}\label{eq:evaluated_c_prime}
\begin{split}
\frac{\partial\langle c\rangle}{\partial t} &+ U(x)\left(1+\frac{\beta^2}{12}\right)\frac{\partial\langle c\rangle}{\partial x}=\\&-D\left(\frac{2\beta^3}{a}+\frac{3}{2}\beta\gamma-\frac{2\beta}{a}\right)\frac{\partial\langle c\rangle}{\partial x}+D\left(1-\frac{1}{12}\frac{U(x)a\beta}{D}+\frac{U(x)^2a^2}{48D^2}-\beta^2\right)\frac{\partial^2\langle c\rangle}{\partial x^2}.
\end{split}
\end{equation}
The detailed derivation of Eq. \ref{eq:evaluated_c_prime} is summarized in Supplementary Section A.

\begin{table}
\small
\caption{Ranked summary of terms in Equation \ref{eq:evaluated_c_prime}.} 
\centering 
\begin{tabular}{ c | c } 
\hline
\multicolumn{2}{l}{\textbf{Left-hand side (LHS)}}\\
\hline
\makecell[l]{$\frac{U_0c_0}{\sigma}\times\mathcal{O}\left(1\right)$} & \makecell[l]{$\frac{\partial \langle c\rangle}{\partial t}$, $U\frac{\partial\langle c\rangle}{\partial x}$}\\
\\[0cm]
\makecell[l]{$\frac{U_0c_0}{\sigma}\times\mathcal{O}\left(\eta Pe_{a_0}\right)$} & \makecell[l]{$\frac{U^2a^2}{48D}\frac{\partial^2\langle c\rangle}{\partial x^2}$}\\
\\[0cm]
\makecell[l]{$\frac{U_0c_0}{\sigma}\times\mathcal{O}\left(\eta\epsilon\right)$} & \makecell[l]{$\frac{1}{12}Ua\beta\frac{\partial^2\langle c\rangle}{\partial x^2}$}\\
\\[0cm]
\makecell[l]{$\frac{U_0c_0}{\sigma}\times\mathcal{O}\left(\epsilon^2\right)$} & \makecell[l]{$\frac{1}{12}\beta^2U\frac{\partial\langle c\rangle}{\partial x}$}\\
\hline
\multicolumn{2}{l}{\textbf{Right-hand side (RHS)}}\\
\hline 
\makecell[l]{$\frac{Dc_0}{\sigma^2}\times\mathcal{O}\left(1\right)$} & \makecell[l]{$D\frac{\partial^2\langle c\rangle}{\partial x^2}$}\\
\\[0cm]
\makecell[l]{$\frac{Dc_0}{\sigma^2}\times\mathcal{O}\left(\frac{\sigma}{\lambda}\right)$} & \makecell[l]{$D\frac{2\beta}{a}\frac{\partial\langle c\rangle}{\partial x}$}\\
\\[0cm]
\makecell[l]{$\frac{Dc_0}{\sigma^2}\times\mathcal{O}\left(\epsilon^2\right)$} & \makecell[l]{$D\beta^2\frac{\partial^2\langle c\rangle}{\partial x^2}$}\\
\\[0cm]
\makecell[l]{$\frac{Dc_0}{\sigma^2}\times\mathcal{O}\left(\frac{c_0^{\prime}}{c_0}\epsilon^2\frac{\sigma}{\lambda}\right)$} & \makecell[l]{$D\frac{2\beta^3}{a}\frac{\partial\langle c\rangle}{\partial x}$, $D\frac{3\beta\gamma}{2}\frac{\partial\langle c\rangle}{\partial x}$}\\
\hline
\end{tabular}
\label{table:eq11_summary} 
\end{table}

Keeping terms on the order of $\frac{Uc_0}{\sigma}\mathcal{O}\left(\eta Pe_{a_0}\right)$ and $\frac{Dc_0}{\sigma^2}\mathcal{O}\left(\frac{\sigma}{\lambda}\right)$ yields the following expression for the development of the area-averaged concentration:
\begin{equation}\label{eq:taylor_final}
\frac{\partial\langle c\rangle}{\partial t}+\left(U(x)-\frac{2\beta D}{a}\right)\frac{\partial\langle c\rangle}{\partial x} = D\left(1+\frac{U(x)^2a^2}{48D^2}\right)\frac{\partial^2\langle c\rangle}{\partial x^2},
\end{equation}
which can also be rearranged and written in this conservative form \citep{Stone2022}:
\begin{equation}\label{eq:taylor_final2}
\frac{\partial\langle c\rangle}{\partial t}+U(x)\frac{\partial\langle c\rangle}{\partial x} = \frac{1}{a^2}\frac{\partial}{\partial x}\left(\left(D+\frac{U(x)^2a^2}{48D}\right)a^2\frac{\partial\langle c\rangle}{\partial x}\right).
\end{equation}
Note that, for $\beta=0$, Equation \ref{eq:taylor_final} reduces to the simple result of Taylor-Aris dispersion for laminar, fully-developed flow within cylindrical channel (with uniform-radius) \citep{Aris1956}.

In the next section, we will use Equation \ref{eq:taylor_final} to develop a description of the development of the axial mean position of the solute and its axial variance. For now, we note the RHS of Equation \ref{eq:taylor_final} contains a prefactor for the second axial derivative $\frac{\partial^2\langle c\rangle}{\partial x^2}$ which has the form of a typical Taylor-Aris dispersion coefficient:
\begin{equation}\label{eq:Deff}
D_{\mathrm{eff}}(x)\equiv D\left(1+\frac{1}{48}\frac{U(x)^2a^2}{D^2}\right)
\end{equation}
In a cylindrical tube with constant radius, the axial variance grows linearly in time according to $2D_{\mathrm{eff}}t$, as $\beta=\gamma=0$ and the mean velocity correction due to $u_r\neq 0$ vanishes. However, we note that the coefficient $D_{\mathrm{eff}}$ is by itself not useful in predicting the time evolution of axial variance when the cross-section is varying in space. This is true even for the simple case of linearly converging or diverging channel (i.e., $\beta=\mathrm{Constant}$). The reason for this is that development of the solute zone variance is a strong function of the advective operator on the left-hand side (LHS). These advective gradients can stretch or shrink the variance as the solute navigates through contractions and expansions, respectively.

\subsection{Time evolution of mean and variance}
We next formulate the problem in terms of analytical expressions of two moments of solute distributions. Note that, despite this use of spatial moments, our analysis does not follow Aris' famous method of moments \citep{Aris1956} as we here focus on the mean and variance of the solute distribution in the typical Taylor-Aris limit where observation times are much longer than the radial diffusion time. Our analysis also uses scaling analyses and approximations for key terms which are not part of the method of moments.

We first define an axial average as
\begin{equation}
    \overline{(\cdot)}\equiv\int_{-\infty}^{\infty}(\cdot)a^2(x^{\prime})\langle c\rangle(x^{\prime}, t)\mathrm{d}x^{\prime}/\int_{-\infty}^{\infty}a^2(x^{\prime})\langle c\rangle(x^{\prime}, t)\mathrm{d}x^{\prime},
\end{equation}
where $x^{\prime}$ is a dummy variable for integration. For more compact notation, we also define $\overline{c}$ as the axial integration of the mean concentration
$\int_{-\infty}^{\infty}a^2(x^{\prime})\langle c\rangle(x^{\prime}, t)\mathrm{d}x^{\prime}\equiv\overline{c}=\mathrm{Constant}$.

To derive the first moment, we multiply Equation \ref{eq:taylor_final} with $x$ and apply the axial average operation $\overline{(\cdot)}$:
\begin{equation}\begin{split}
\int_{-\infty}^{\infty}xa^2\frac{\partial\langle c\rangle}{\partial t}\mathrm{d}x+\int_{-\infty}^{\infty}xUa^2\frac{\partial\langle c\rangle}{\partial x}\mathrm{d}x - \int_{-\infty}^{\infty}xa^2\frac{2\beta D}{a}\frac{\partial\langle c\rangle}{\partial x}\mathrm{d}x \\
=\int_{-\infty}^{\infty}xa^2D\left(1+\frac{1}{48}\frac{U^2a^2}{D^2}\right)\frac{\partial^2\langle c\rangle}{\partial x^2}\mathrm{d}x
\end{split}\end{equation}
We evaluate integrals using integration by parts, divide both sides with $\overline{c}$, and arrive at an ODE for the axial mean position $\overline{x}(t)$ as follows:
\begin{equation}\label{eq:first_moment}
\begin{split}
\frac{\mathrm{d}\overline{x}}{\mathrm{d}t}=U_0a_0^2\overline{\left[\frac{1}{a^2}\right]}+2D\overline{\left[\frac{\beta}{a}\right]}
\end{split}
\end{equation}
The two terms on the RHS are expressions of the form $\overline{f(x)}$. Some analysis of this term is essential for further analytical simplification of the problem. We consider a Taylor expansion of $f(x)$ about the mean axial position $\overline{x}$ as follows:
\begin{equation}\label{eq:taylor_fx}
f(x)=f(\overline{x})+(x-\overline{x})f'(\overline{x})+\frac{1}{2}(x-\overline{x})^2f''(\overline{x})+\frac{1}{6}(x-\overline{x})^3f'''(\overline{x})+...
\end{equation}
To simplify the expressions, we define $\widetilde{f}\equiv f(\overline{x})$. Insert the expression in Eq. \ref{eq:taylor_fx} into Eq. \ref{eq:first_moment} and keep the dominant term, we derive the following ODE for $\overline{x}(t)$:
\begin{equation}\label{eq:x_bar}
\frac{\mathrm{d}\overline{x}}{\mathrm{d}t}= U(\overline{x})+2D\widetilde{\left[\frac{\beta}{a}\right]}+\mathcal{O}\left(U_0\left(\frac{\sigma}{\lambda}\right)^2, \frac{D}{\lambda}\left(\frac{\sigma}{\lambda}\right)^2\right),
\end{equation}
which reduces to the standard Taylor result for $\beta(x)=0$.

To derive the second moment equation of Equation \ref{eq:taylor_final}, we multiply Equation \ref{eq:taylor_final} by $(x-\overline{x})^2$, apply the axial average operation, perform integration by parts, and divide both sides by $\overline{c}$. This then yields the following ODE for the dynamics of $\sigma_x^2$:
\begin{equation}\label{eq:second_moment}
\begin{split}
\frac{\mathrm{d}\sigma^2_x}{\mathrm{d}t}&=2Ua^2\left[\overline{\frac{x}{a^2}}-\overline{x}\overline{\frac{1}{a^2}}\right]+\frac{U^2a^4}{24D}\overline{\frac{1}{a^2}}+2D+4D\overline{\frac{\beta}{a}(x-\overline{x})}
\end{split}
\end{equation}
Note that the RHS of Equation \ref{eq:second_moment} includes expressions of the form $\overline{f(x)}$, $\overline{(x-\overline{x})f(x)}$, and $\overline{xf(x)}$. We again simplify the latter second and third terms using Taylor expansion at the mean axial position $\overline{x}$ as follows:
\begin{equation}\label{eq:taylor_xfx}
\begin{split}
(x-\overline{x})f(x)&=(x-\overline{x})f(\overline{x})+(x-\overline{x})^2f'(\overline{x})+\frac{1}{2}(x-\overline{x})^3f''(\overline{x})+\frac{1}{6}(x-\overline{x})^4f'''(\overline{x})...\\
xf(x)&=\overline{x}f(\overline{x}) + (x-\overline{x})(\overline{x}f^{\prime}(\overline{x})+f(\overline{x}))+(x-\overline{x})^2(\frac{1}{2}\overline{x}f^{\prime\prime}(\overline{x})+f^{\prime}(\overline{x}))\\
&+\frac{1}{6}(x-\overline{x})^3(\overline{x}f'''(\overline{x})+3f''(\overline{x}))+...
\end{split}
\end{equation}
We insert the expression in Eq. \ref{eq:taylor_xfx} into Eq. \ref{eq:second_moment}, keep the term of order $D\times\mathcal{O}\left(1\right)$, $D\times\mathcal{O}\left(Pe_{a_0}^2\right)$, $D\times\mathcal{O}\left(Pe_{a_0}\frac{\sigma^2}{a_0\lambda}\right)$, and arrive at an ODE of the axial variance $\sigma_x^2(t)$ as
\begin{equation}\label{eq:var}
\begin{split}
\frac{\mathrm{d}\sigma^2_x}{\mathrm{d}t}&=2D+\frac{U^2a^4}{24D}\widetilde{\left[\frac{1}{a^2}\right]} - 4Ua^2\sigma^2_x\widetilde{\left[\frac{\beta}{a^3}\right]}+D\mathcal{O}\left(\left(\frac{\sigma}{\lambda}\right)^2, Pe_{a_0}^2\left(\frac{\sigma}{\lambda}\right)^2, Pe_{a_0}\frac{\sigma^2}{a_0\lambda}\frac{\sigma}{\lambda}\mathrm{Skew}_x \right),
\end{split}
\end{equation}
In the latter equation, as $\sigma/\lambda$ approaches unity, the dominant error term is $D\times\mathcal{O}\left(Pe_{a_0}\frac{\sigma^2}{a_0\lambda}\frac{\sigma}{\lambda}\mathrm{Skew}_x \right)$ term. For $\sigma/\lambda$ approaching unity, this has the following order of magnitude:
\begin{equation}\label{eq:dominant_error}
2U_0a_0^2\mathrm{Skew}_x\sigma_x^3\widetilde{\frac{3\beta^2-a\gamma}{a^4}},
\end{equation}
Here $\mathrm{Skew}_x$ is the axial skewness of the solute zone, defined as $\mathrm{Skew}_x=\overline{(x-\overline{x})^3}/\sigma_x^3$. Detailed derivation of Eq. \ref{eq:dominant_error} is summarized in Supplementary Section B. Note that Equation \ref{eq:var} also reduces to standard Taylor results when $\beta(x)=0$. Equation \ref{eq:x_bar} and \ref{eq:var} are two coupled ODEs that predict the time evolution of axial mean and variance of solute based on channel geometry and the assumption of lubrication theory in velocity field. In Section 3, we will use these two equations to predict the time evolution of the axial mean and variance of solute zone for various interesting channel geometries.

\subsection{Analytical derivation of the boundary between the regimes of transient positive and negative growth of axial variance}
\begin{figure}
\centerline{\includegraphics[width=1.0\textwidth]{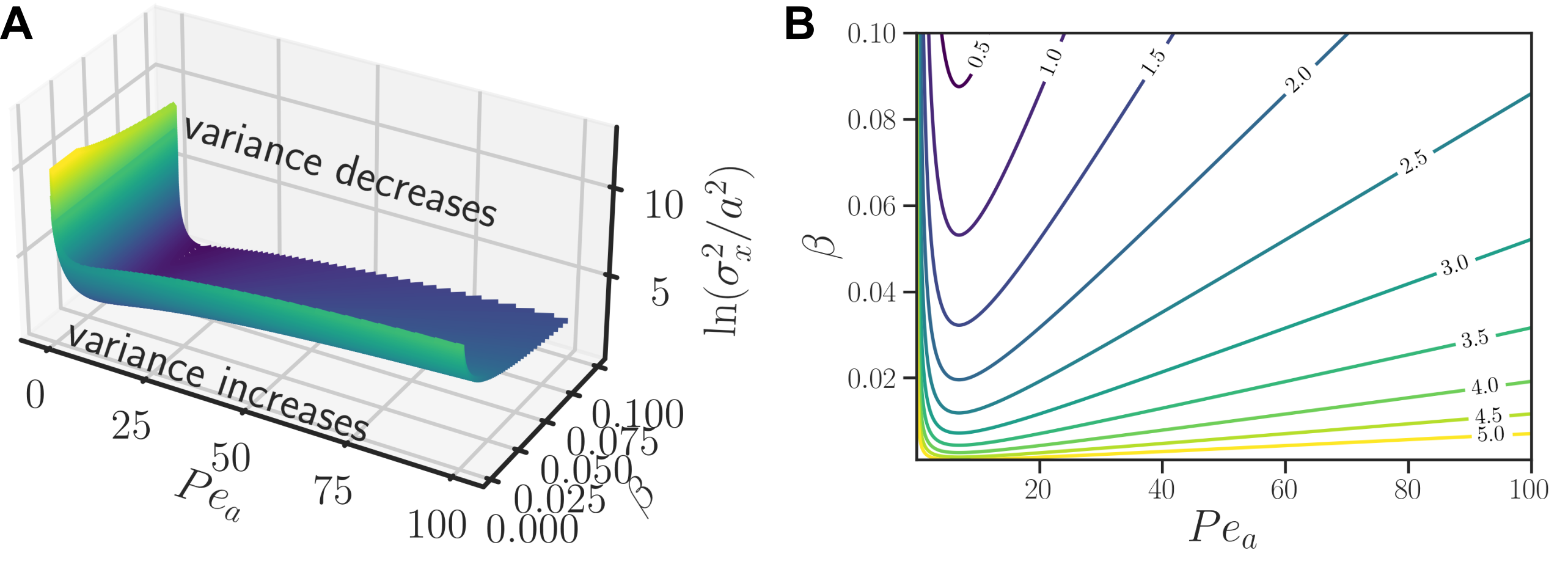}}
\caption{Regimes of transient positive and negative growth of axial variance of solute. (A) Surface of zero variance growth in a space of Peclet number, $\beta$, and the natural log of the local ratio between variance and squared radius ($\ln(\sigma_x^2/a^2)$). This surface is computed analytically from Equation \ref{eq:regime}. Shown is a surface where the axial variance rate of growth, $\frac{\mathrm{d}\sigma_x^2}{\mathrm{d}t}$, is approximately 0. (B) The contour plot shows the horizontal cross-section curves of the zero transient axial variance growth surface in a space of Peclet number, $\beta$, and $\ln(\sigma_x^2/a^2)$, at different $\ln(\sigma_x^2/a^2)$ values (labeled on each contour line).
}\label{fig:regime}
\end{figure}

Advective dispersion can cause the axial variance of the solute zone to decrease as the solute travels through a region where the channel is expanding. Such an expanding channel region can be, qualitatively, characterized by a positive and sufficiently large value of $\beta$ and a sufficiently large $Pe_a$ (so that molecular diffusion does not completely overpower the advective effect), and a sufficiently large value of the axial variance relative to the local channel radius. We here explore this effect. Enabled by our analytical approach, we will formulate the boundary between the physical regimes of transient positive and negative growth of axial variance of the solute. Rearranging terms in Eq. \ref{eq:var}, we see that
\begin{equation}\label{eq:regime}
\frac{1}{D}\frac{\mathrm{d}\sigma^2_x}{\mathrm{d}t}=2\left(1+\frac{1}{48}Pe_a^2\right)-4Pe_a\beta\left(\frac{\sigma_x^2}{a^2}\right).
\end{equation}
This expression yields a useful description of the boundary between the regimes of transient positive and transient negative growth of axial variance in terms of $\beta$, $Pe_a$, and $\sigma_x^2/a^2$. Note that here the negative growth of variance is a transient phenomenon, and is limited simply to the axial width of the sample. The growth of the three-dimensional extent of the solute cloud can, of course, never be negative. 
Figure \ref{fig:regime} shows a plot of this boundary as a three-dimensional surface in terms of the aforementioned three variables. That is, the surface (Fig. \ref{fig:regime}A) delineates the regions of transient positive and negative growth of axial variance as a function of local Peclet number ($Pe_a$), local slope of channel radius distribution ($\beta$), and local ratio between variance and squared radius ($\sigma_x^2/a^2$). Figure \ref{fig:regime}B shows the contours of horizontal cross-section of the surface for various values of $\ln(\sigma_x^2/a^2)$(labeled on each contour line). Points above the surface exhibit a transient negative rate of axial variance growth, while points below have a positive rate of growth. Accordingly, the surface defines the solution for transient zero growth in variance in this three-parameter space. As expected, the surface asymptotes toward the $Pe_a$-$\beta$ plane for finite $Pe_a$ and larger values of $\beta$. We also see that for fixed and finite values of (positive) $\beta$, setting the LHS of Eq. \ref{eq:regime} to zero reveals the critical variance value $\sigma_x^2/a^2$ required for transient negative growth of axial variance as follows:
\begin{equation}
\frac{\sigma_x^2}{a^2}=\frac{1}{4\beta}\left(\frac{2}{Pe_a}+\frac{Pe_a}{24}\right)
\end{equation}
We now differentiate the latter expression with respect to $Pe_a$, set it equal to zero, and arrive at the minimal variance required for transient negative axial variance growth:
\begin{equation}
\min\left(\frac{\sigma_x^2}{a^2}\right)=\frac{1}{4\sqrt{3}\beta}.
\end{equation}
The minimum variance occurs at $Pe_a=\sqrt{48}$.

\subsection{Approximations useful for channel design}
The analytical expression of Equation \ref{eq:regime} is useful in engineering channel geometries which will generate desired variance evolutions in the Taylor-Aris dispersion regime. We can simplify such a solution further by implementing the following approximation:
\begin{equation}
\begin{split}
    \frac{\mathrm{d}}{\mathrm{d}t}&\approx\frac{\mathrm{d}\overline{x}}{\mathrm{d}t}\frac{\mathrm{d}}{\mathrm{d}x}
\end{split}
\end{equation}
Inserting this approximation into Equation \ref{eq:var}, and substituting the definition of $\beta$, we obtain an ODE for $a(x)$ given a desired spatial distribution $\sigma_x^2(x)$:
\begin{equation}\label{eq:arb_var}
    \frac{\mathrm{d}a}{\mathrm{d}x}\approx\frac{-\frac{\mathrm{d}\sigma_x^2}{\mathrm{d}x}+\frac{2a^2D}{U_0a_0^2}+\frac{U_0a_0^2}{24D}}{\frac{4\sigma_x^2}{a}+\left(\frac{2D}{U_0a_0^2}\frac{\mathrm{d}\sigma_x^2}{\mathrm{d}x}\right)a}
\end{equation}
We can then numerically solve Equation \ref{eq:arb_var} to obtain the shape of the engineered channel. In Section 3.4, We will use this approximation to design channels that can result in specific variance evolution pattern in space.

\subsection{Brownian Dynamics simulations}
We used Brownian dynamics simulations to benchmark and evaluate our analytical expressions for variance evolution. Each Brownian dynamics simulation consisted of tracking 5,000 point-particles which were initially uniformly distributed along the radius and normally distributed along the streamwise direction. We considered a mean axial position of zero (by definition) and an initial axial variance of $\sigma_{x,0}^2$. We set $\sigma_{x,0}^2 = 300a_0^2$ for the two engineered channels in Section 3.4, and we set $\sigma_{x,0}^2 = 10a_0^2$ for all other cases. The velocity field follows Equation \ref{eq:lubrication}. The system evolves according to the forward Euler method.

For the time steps, we set the ratio between diffusion time scale ($a_0^2/D$) to time step of discretization to 40. Each reflection at the wall was checked twice at each time step. Unless specified, the (constant, initial) Peclet number based on initial radius ($Pe_{a_0}$) was 10. For the three periodic channels shown in Section 3.2, the initial Peclet number was 0.1, 1, 10, 100, and 1000. Five simulations from different random seed initial conditions were computed for each case, and the reported values are their average. The numerical computation of axial averaged quantities $\overline{f(x)}$ is computed with $\overline{f(x)}=\sum_{i=1}^{N_{\mathrm{particle}}}f(x_i)/N_{\mathrm{particle}}$, where $x_i$ is the axial location of the $i$-th particle. The program was written in Python 3 and is available on \href{https://github.com/jrchang612/Taylor_dispersion_arbitrary}{Github} (jrchang612/Taylor\_dispersion\_arbitrary). The detailed parameters used for the simulation presented in the figures are summarized in Supplementary Section C.

\section{Results and Discussion}
\subsection{Diverging and converging conical channels}
\begin{figure}
\centerline{\includegraphics[width=1.0\textwidth]{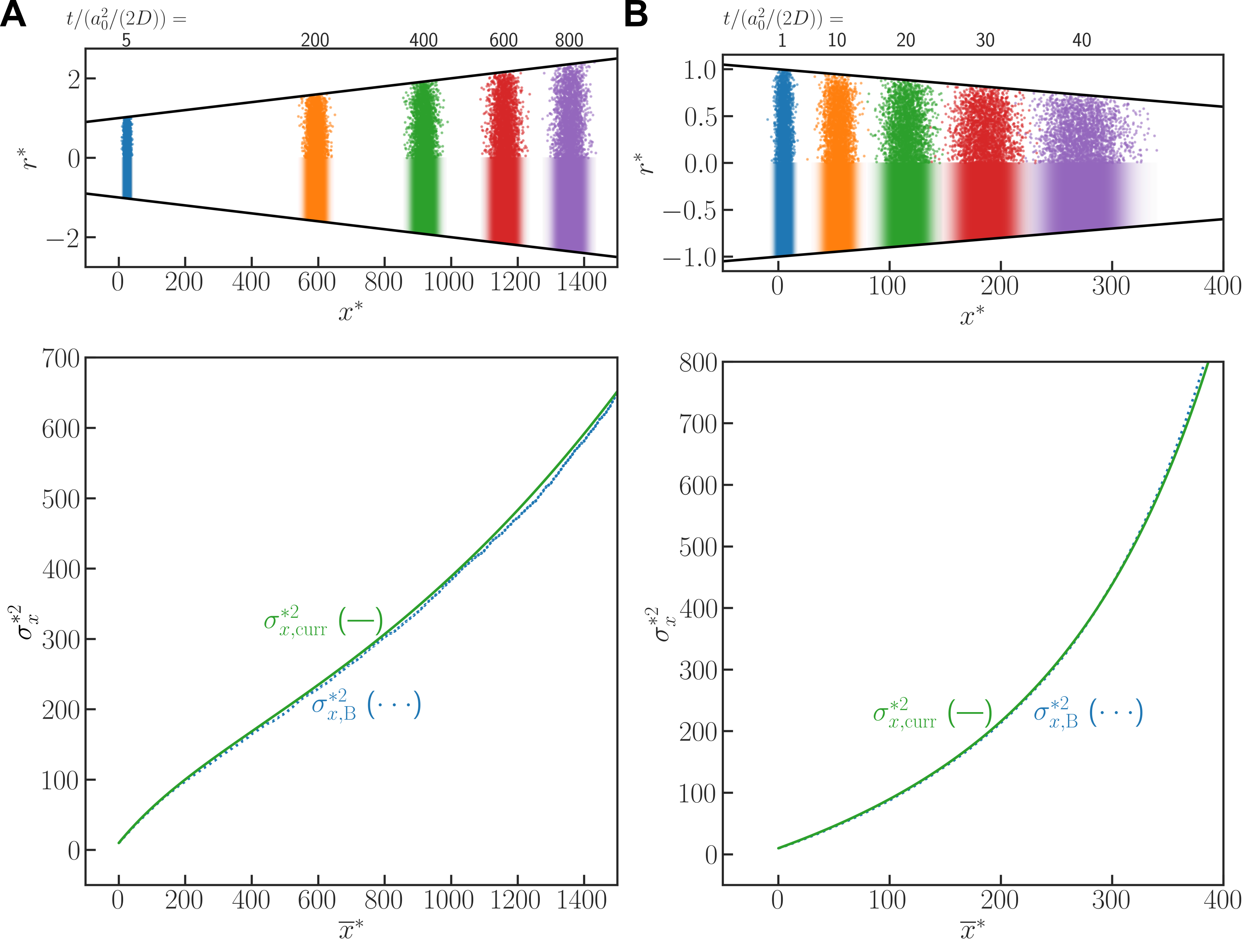}}
\caption{Taylor-Aris dispersion in (A) diverging and (B) converging conical channels. The plots in the top row show results from Brownian dynamic simulation (upper half) along with the predicted axial distribution of area-averaged concentration (bottom half). The bottom two plots show solute axial variance as a function of $\overline{x}^*$. Plotted are axial variance from the current analytical model (subscript "curr", Equation \ref{eq:var}) and from Brownian dynamics simulation (subscript "B"). Axial variance computed using Equation \ref{eq:var} shows excellent agreement with Brownian dynamics simulations. 
}\label{fig:divconv}
\end{figure}

We first apply our analysis to a simple case of diverging and converging conical section channels. Figures \ref{fig:divconv}A and \ref{fig:divconv}B shows solutions for dispersion of a solute zone in linearly diverging and converging channels, respectively. The top two figures in Figure \ref{fig:divconv} show particle zones at five non-dimensional times $2Dt/a_0^2$ (from 5 to 800) as they migrate through the channels. The axes are non-dimensionalized coordinates $r^*=r/a_0$ and $x^*=x/a_0$. Note the intensely exaggerated height-to-length aspect ratio of the figures, chosen here for clarity of presentation. The top half of the solute concentration in the channel are raw data results of the three-dimensional Brownian dynamics simulations. The bottom half of the channels show the (one-dimensional) area-averaged axial distribution of concentration of solute from the model developed here (i.e. numerical solutions of ODEs given by Eqs.\ref{eq:x_bar} and \ref{eq:var}). Note that we present and plot the predicted axial distribution assuming a Gaussian distribution as we only have the information from mean and variance. However, the derivation of Eqs.\ref{eq:x_bar} and \ref{eq:var} do not rely on a Gaussian distribution assumption. The solute is initially uniformly distributed over the cross-sectional area of the channel, and the initial standard deviation width of the axial solute distribution are each set equal to the same value (equal to $\sqrt{10}a_0=\sqrt{10}a(x=0)$ of the diverging channel). The current model has very good comparison with the Brownian dynamics simulations. The characteristic $Pe_a$ values here are order $\mathcal{O}(10)$ and this results in negligible radial gradients of particle density in the Brownian dynamics. Hence, the particle clouds from Brownian dynamics are fairly symmetric about $x^*$. Note the general trend of relatively rapid increase of zone variance in the converging case. By comparison, the expansion in the diverging channel limits the growth of variance. For this case, the divergence is not sufficiently pronounced to result in transient negative growth of axial variance (but we shall explore such cases later below.)

The bottom two plots in Figure \ref{fig:divconv} are plots of the axial variance of the area-averaged axial distribution of solute concentration from the current model ($\sigma_{x,\mathrm{curr}}^{*2}$) and  Brownian dynamics ($\sigma_{x,\mathrm{B}}^{*2}$). These quantities are plotted as a function of the axial mean position of solute zone, $\overline{x}^*$. There is excellent agreement between the current model and the Brownian dynamics simulation for both the diverging and converging cases, demonstrating the ability of the current model to predict the spatial evolution pattern of axial variance. 

\subsection{Periodic channels}
\begin{figure}
\centerline{\includegraphics[width=1.0\textwidth]{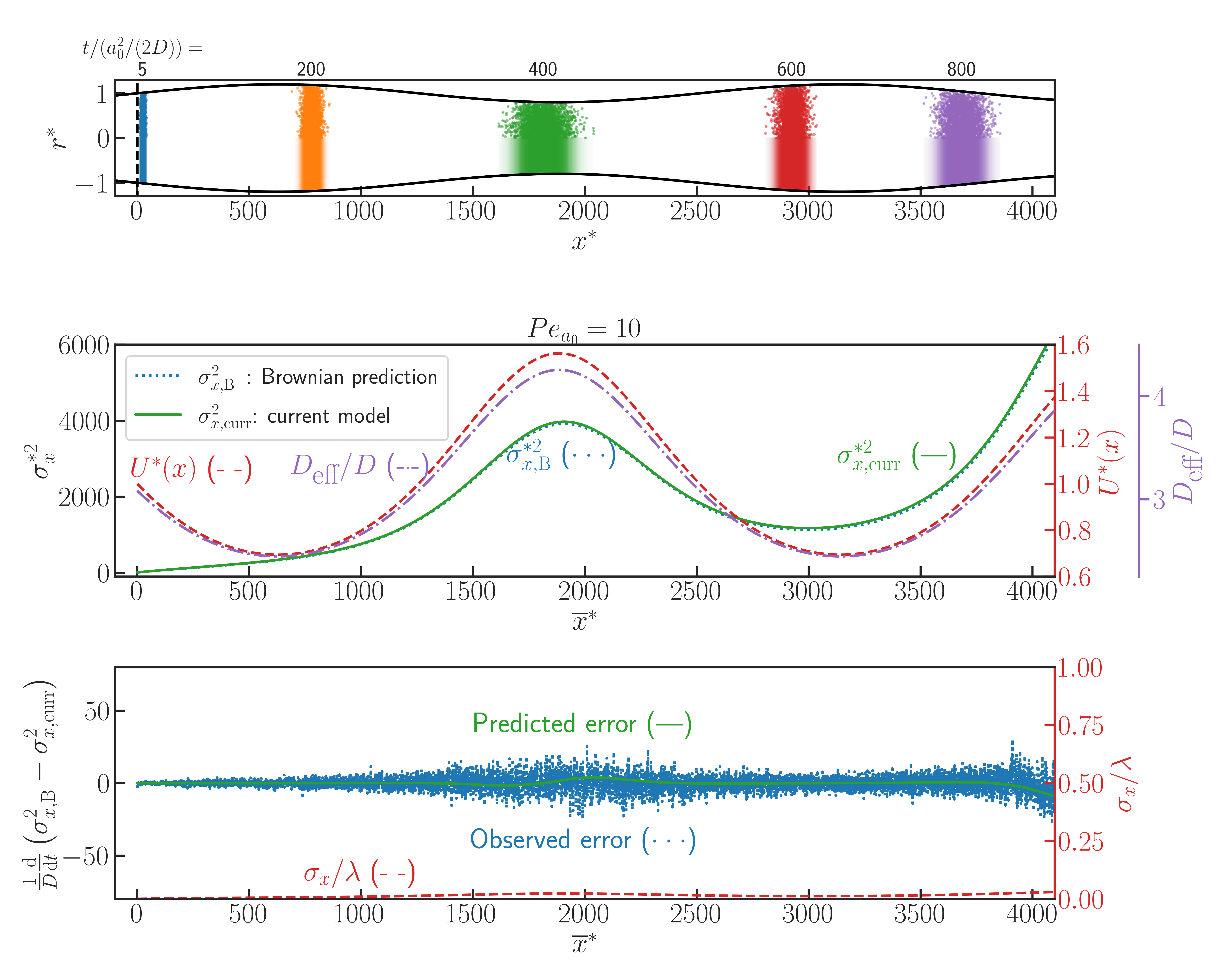}}
\caption{Taylor-Aris dispersion in a channel with a sinusoidal (periodic) radius distribution, $a(x)$. The top plot shows results from a Brownian dynamic simulation (upper half) and axial solute distribution predicted with the current model (Equation \ref{eq:var}). The middle plot shows the distribution of $\sigma_{x,\mathrm{B}}^{*2}$, $\sigma_{x,\mathrm{curr}}^{*2}$, $U^*(x)$, and $D_{\mathrm{eff}}/D$, each as a function of the non-dimensional axial location along the channel, $\overline{x}/a_0$. $U^*(x)$ and $D_{\mathrm{eff}}^*$ are periodic functions and shown as a reference. The bottom plot shows the predicted (Eq. \ref{eq:dominant_error}) and observed error of our current model in the growth rate of axial variance. Also shown for reference is the ratio between the square root of the axial variance and channel wavelength $\sigma_x/\lambda$. Variance computed using Equation \ref{eq:var} shows excellent agreement with Brownian dynamics simulations. Note variance averaged along the axial spatial period increases monotonically as expected. The error in axial variance growth rate is small and matches the predicted error from Eq. \ref{eq:dominant_error}. This shows the accuracy of our model when $\sigma_x/\lambda$ is small.
}\label{fig:periodic}
\end{figure}

We next apply our model to a periodic channel with sinusoidal radius distribution. Similar to the top plots of Figure \ref{fig:divconv}, the top plot of Figure \ref{fig:periodic} shows a plot of the channel geometry in coordinates of $r^*$ versus $x^*$ in a highly exaggerated aspect ratio for clarity. We show results for $Pe_{a_0}$ of 10 where $a_0$ is defined as the radius at $x=0$ and is also the axially averaged radius. Again, the channel shows plots of solute zones at non-dimensional times ranging from 5 to 800. The top half of the solute zones are raw results from the Brownian dynamics simulations, while the bottom half are plots of the predicted axial distribution of area-averaged concentration from the current model. Again, we see agreement between the axial distributions of Brownian dynamics particle densities and the axial distributions from the model. Note how the model captures the strongly positive and negative growth of axial variance as the solute traverses through contractions and expansions, respectively. For example, note the rapid increases in axial variance due to the contraction just upstream of $x^*=1800$, and then the subsequent rapid decrease in axial variance caused by the advective effects of the solute expanding from $x^*=1800$ to just upstream of $x^*=3000$. Note that the particle clouds from Brownian dynamics are also fairly symmetric about $x^*$. 

The middle plot of Figure \ref{fig:periodic} shows the axial variance of the solute zone as a function of solute average axial location $\overline{x}^*$. Plotted are the axial variance from the Brownian dynamics simulation and the current model. Note the excellent agreement between these two models showing how the current model captures very well the detailed development of the axial variance. Note also how the axial variance at equal values of the phase increases and the axial variance averaged over the period increases monotonically. At the end of the current section, we will consider this development further and use our model to analyze period-averaged axial variance over long times.

Figure \ref{fig:periodic} also shows the non-dimensional area-averaged flow velocity and the local normalized effective dispersion coefficient, $D_{\mathrm{eff}}/D$ (c.f. two ordinances on the RHS of the plot). Both $U$ and $D_{\mathrm{eff}}^*$ are periodic functions, antiphase to the shape of the channel. $D_{\mathrm{eff}}^*$ increases as the channel contracts and decreases as the channel expands, as we expected. However, even when the variance is actually decreasing (e.g. between $x^* = 2000$ to $3000$) the effective dispersion coefficient $D_{\mathrm{eff}}/D$ remains positive and greater than unity. This demonstrates the inability of effective dispersion coefficient alone to describe the axial variance evolution (including here in a periodic channel).

The bottom plot of Figure \ref{fig:periodic} shows the observed and predicted error in dimensionless growth rate of axial variance of solute as a function of $\overline{x}^*$. Also shown in the figure is the ratio between the square root of axial variance and channel wavelength $\sigma_x/\lambda$. Consistent with the middle plot of Figure \ref{fig:periodic}, the observed error in axial variance growth rate is small and stochastic, and it matches with the predicted error from Eq. \ref{eq:dominant_error}. The ratio $\sigma_x/\lambda$ is much smaller than 1 throughout the simulation, confirming the accuracy of our model when the assumptions listed in Table \ref{table:assumptions_summary} are satisfied.

\begin{figure}
\centerline{\includegraphics[width=1.0\textwidth]{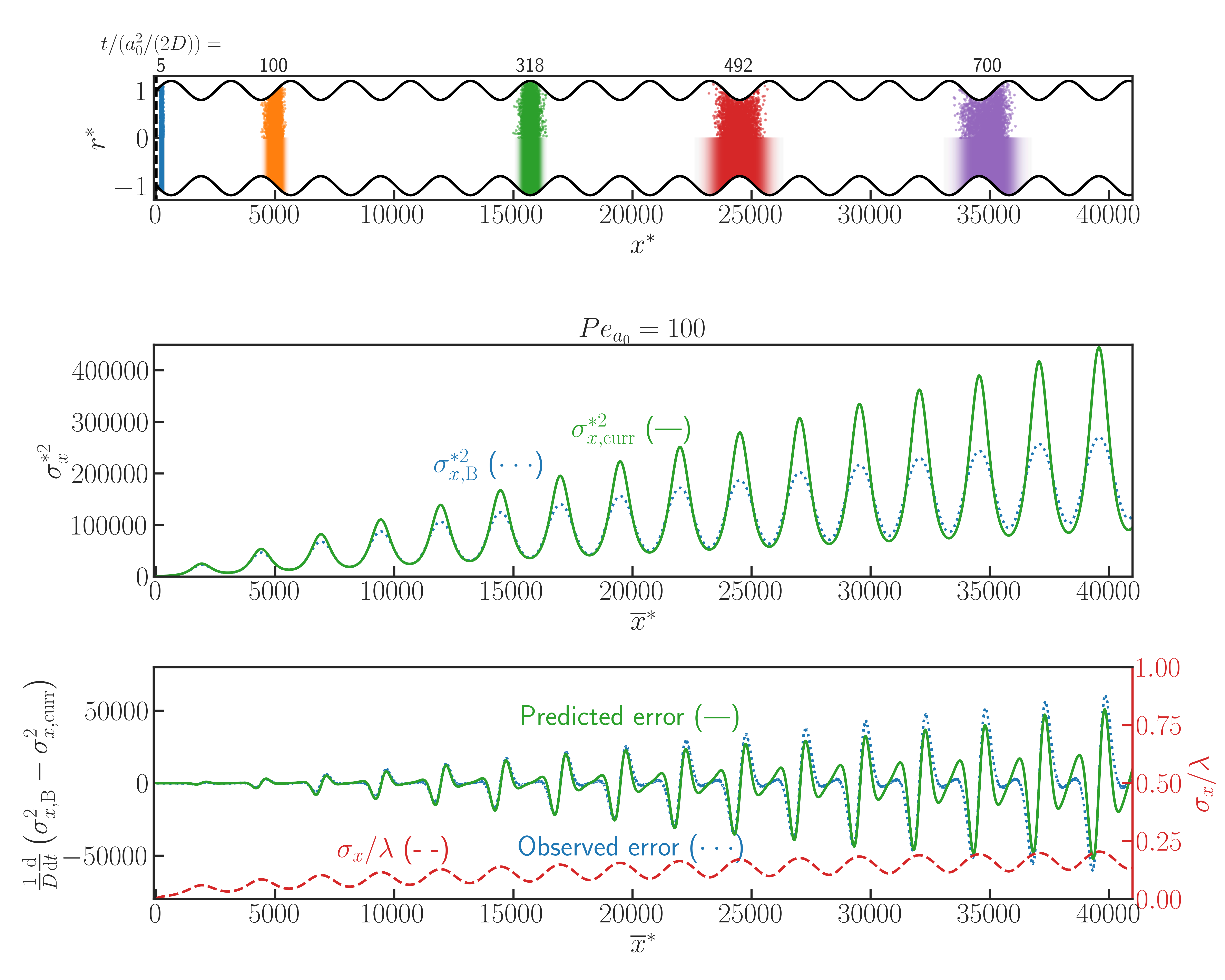}}
\caption{Taylor-Aris dispersion in a channel with a sinusoidal (periodic) radius distribution, $a(x)$. The channel is the same as the one in Figure \ref{fig:periodic}, but the initial Peclet number $Pe_{a_{0}}$ is 100. The top plot shows results from a Brownian dynamic simulation (upper half) and axial solute distribution predicted with the current model (Eq. \ref{eq:var}). The middle plot shows distribution of $\sigma_{x,\mathrm{B}}^{*2}$ and $\sigma_{x,\mathrm{curr}}^{*2}$, each as a function of the non-dimensional axial location along the channel, $\overline{x}/a_0$. The bottom plot shows the predicted (Eq. \ref{eq:dominant_error}) and observed error of our current model in the growth rate of axial variance. Also shown for reference is the ratio between the square root of axial variance and channel wavelength $\sigma_x/\lambda$. Variance computed using Equation \ref{eq:var} shows excellent agreement with Brownian dynamics simulations when the channel is expanding, but overestimates the axial variance when the channel is contracting. Note variance averaged along the axial spatial period increases monotonically as expected. The predicted error in axial variance growth rate using Equation \ref{eq:dominant_error} shows excellent agreement with the observed error, especially when $\sigma_x/\lambda$ is small. As $\sigma_x/\lambda$ increases, there is more deviation between observed and predicted error in axial variance growth rate.
}\label{fig:periodic_Pe100}
\end{figure}

We next increase the Peclet number of the system to evaluate how the model assumptions become inaccurate and the current model fails. We repeat our simulation in the same channel as presented in Figure \ref{fig:periodic} but increase the initial Peclet number to 100. Similar to the top plots of Figure \ref{fig:periodic}, the top plot of Figure \ref{fig:periodic_Pe100} shows a plot of the channel geometry in coordinates of $r^*$ versus $x^*$ in a highly exaggerated aspect ratio for clarity. Again, the channel shows plots of solute zones at non-dimensional times ranging from 5 to 700, with non-dimensional time points at 318 and 492 to demonstrate the best and worst-case scenarios in predictions. The top half of the solute zones are raw results from the Brownian dynamics simulations, while the bottom half are plots of the predicted axial distribution of area-averaged concentration from the current model. Again, we see overall agreement between the axial distributions of Brownian dynamics particle densities and the axial distributions from the model. The model captures the strongly positive and transient negative growth of axial variance as the solute traverses through contractions and expansions, respectively. Our model tends to overestimate the fluctuation in axial variance, especially when the solute transverses through contractions. This overestimation is likely due to the fact that the width of the solute zone becomes comparable to the wavelength of the channel. For example, when the solute mean position is within a contraction, the leading and trailing ends of the solute zone are in expansion zones.

The middle plot of Figure \ref{fig:periodic_Pe100} shows the axial variance of the solute zone as a function of $\overline{x}^*$. Plotted are the variance from the Brownian dynamics simulation and the current model. There is a good agreement between these two models in trends of development of the variance, and an excellent agreement when $\overline{x}^*$ is less than 5000. The agreement between the two models remains very good when the solute passes through expansions in the channel, but the current model overestimates the axial variance when the solute passes through contraction. 

The bottom plot of Figure \ref{fig:periodic_Pe100} again shows the observed and predicted error in dimensionless growth rate of axial variance of solute as a function of $\overline{x}^*$. Also shown in the figure is the ratio between the square root of axial variance and channel wavelength $\sigma_x/\lambda$. The magnitude of observed error grows as the axial variance grows, and the error is most pronounced when the solute transverses through contractions. There is an excellent agreement between the predicted error from Eq. \ref{eq:dominant_error} when $\overline{x}^*$ is less than 20000, but there is more deviation between the two predictions as $\sigma_x/\lambda$ grows larger.

\begin{figure}
\centerline{\includegraphics[width=1.0\textwidth]{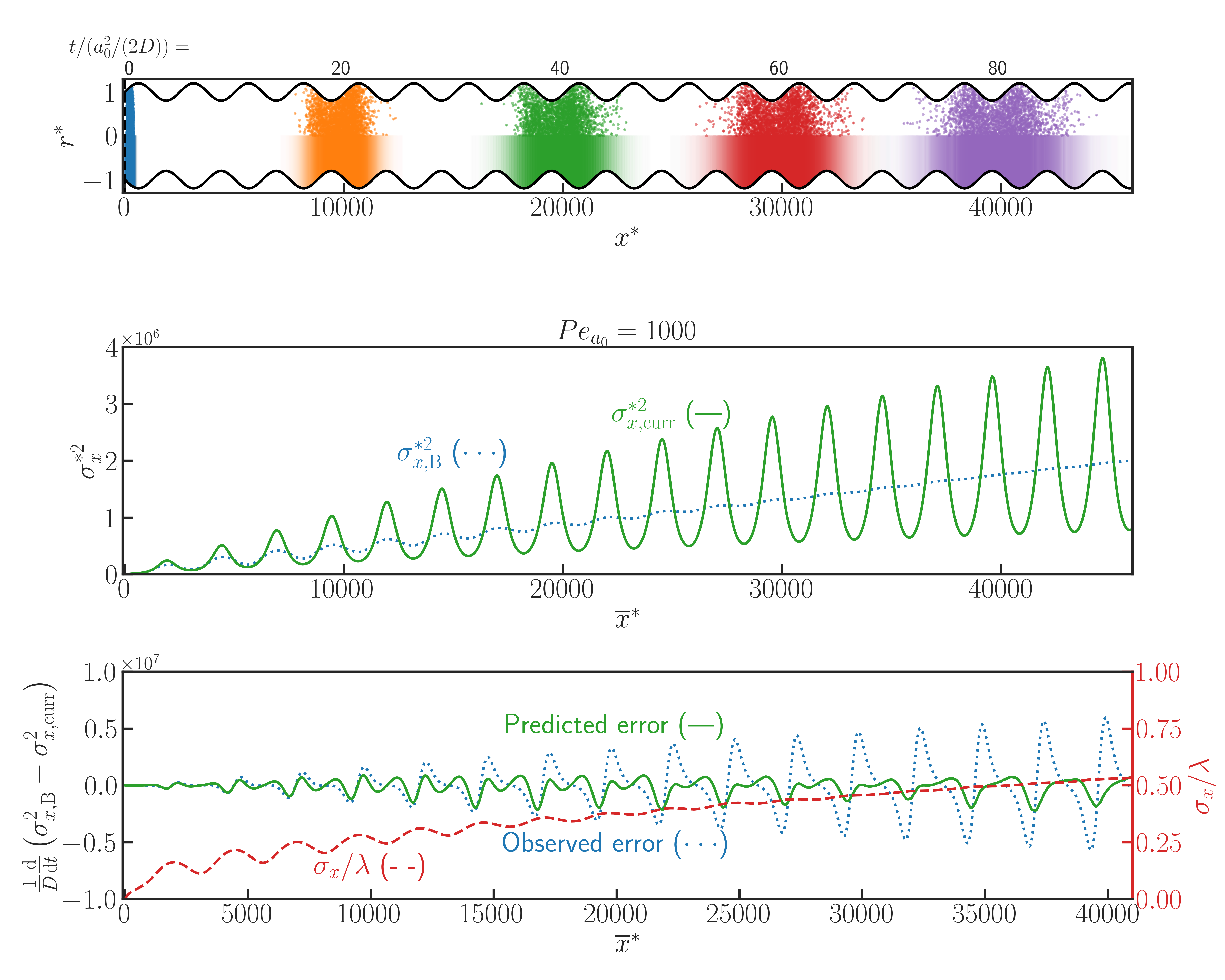}}
\caption{Taylor-Aris dispersion in a channel with a sinusoidal (periodic) radius distribution, $a(x)$. The channel is the same as the one in Figure \ref{fig:periodic}, but the initial Peclet number $Pe_{a_{0}}$ is 1000. The top plot shows results from a Brownian dynamic simulation (upper half) and axial solute distribution predicted with the current model (Eq. \ref{eq:var}). The middle plot shows distribution of $\sigma_{x,\mathrm{B}}^{*2}$ and $\sigma_{x,\mathrm{curr}}^{*2}$, each as a function of the non-dimensional axial location along the channel, $\overline{x}/a_0$. The bottom plot shows the predicted (Eq. \ref{eq:dominant_error}) and observed error of our current model in the growth rate of axial variance. Also shown for reference is the ratio between the square root of axial variance and channel wavelength $\sigma_x/\lambda$. Variance computed using Equation \ref{eq:var} only shows agreement with Brownian dynamics simulations in the very beginning. Our model fails to predict the monotonic growth of axial variance when the axial variance on the same order as the channel wavelength. The predicted error in axial variance growth rate using Equation \ref{eq:dominant_error} shows fair agreement with the observed error for $\overline{x}^{*}$ below 10000. As the ratio between axial variance and channel wavelength approaches unity, the predicted error also become inaccurate as the basic assumptions of our model are no longer valid.
}\label{fig:periodic_Pe1000}
\end{figure}

We further increases the initial Peclet number of the system to 1000 using the same channel of Figure \ref{fig:periodic} and Figure \ref{fig:periodic_Pe100}. Similar to the topmost plots of Figure \ref{fig:periodic_Pe100}, the top of Figure \ref{fig:periodic_Pe1000} shows a plot of the channel geometry in coordinates of $r^*$ versus $x^*$ in a highly exaggerated aspect ratio for clarity. Again, the shown are solute zones at non-dimensional times ranging from 5 to 80. The top half of the solute zones are raw results from the Brownian dynamics simulations, while the bottom half are Gaussian distributions with mean and axial variances predicted by the current model. The agreement between the two is now merely qualitative. Note that, for these conditions, the sample zone axial width quickly becomes on the same order as the wavelength of the channel.

The middle plot of Figure \ref{fig:periodic_Pe1000} shows the axial variance of the solute zone as a function of $\overline{x}^*$. Plotted are the variance from the Brownian dynamics simulation and the current model. The agreement between the two is limited to the initial development of the axial variance ($\overline{x}^*<2000$). The axial variance growth in Brownian dynamics simulation becomes monotonic after $\overline{x}^*>30000$, but the current model is not able to capture this.

\begin{figure}
\centerline{\includegraphics[width=1.0\textwidth]{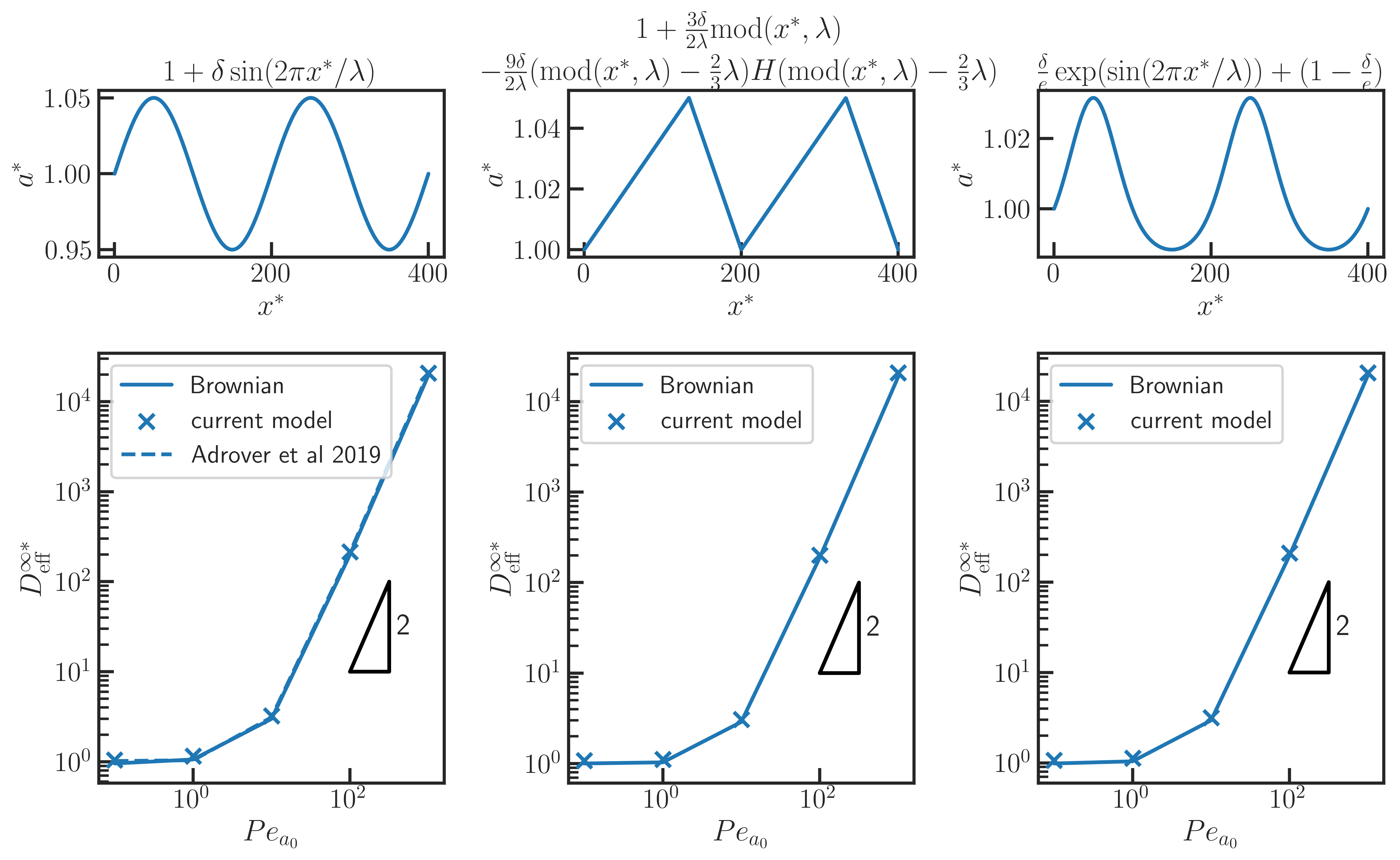}}
\caption{Normalized long-term effective dispersion coefficient $D_{\mathrm{eff}}^{\infty *}$ as a function of $Pe_{a_0}$ for three different periodic channels with equal period and similar radius amplitude ($\delta=0.05$, $\lambda=200$). All plots show the $D_{\mathrm{eff}}^{\infty *}$ computed using Eq.\ref{eq:var} and with a Brownian dynamics simulation. The plot on the left also shows a comparison with the expression derived by Adrover et al. (2019) for a sinusoidal channel \citep{Adrover2019}. Note the plots across all three channels are very similar (but not exactly the same) in magnitude and shape. This similarity reflects the fact that the long-term development of the solute in periodic channels is most strongly a function of channel amplitude and a weak function of channel shape.
}\label{fig:literature}
\end{figure}

The bottom plot of Figure \ref{fig:periodic_Pe1000} again shows the observed and predicted error in dimensionless growth rate of axial variance of solute as a function of $\overline{x}^*$. Also shown in the figure is the ratio between the square root of axial variance and channel wavelength $\sigma_x/\lambda$. The magnitude of observed error grows as the axial variance grows, and the error is now pronounced in both channel contractions and expansions. The predicted error from Eq. \ref{eq:dominant_error} can only capture the observed error in the initial development of axial variance ($\overline{x}^*<10000$), but becomes inaccurate as $\sigma_x/\lambda$ grows above 0.3. This finding suggests that the accuracy of our model depends less on the Peclet number itself but more sensitive to the ratio $\sigma_x/\lambda$. At large Peclet number, since the axial variance growth rate is much faster (as discussed in Figure \ref{fig:literature}) and $\sigma_x/\lambda$ quickly approaches and exceeds unity, the current model becomes inaccurate. A zoomed-in comparison of the solute zone subject to Peclet number of 10, 100 and 1000 is in Figure S1.

We next consider the long-term averaged dispersion coefficient in periodic channels. As with previous studies of dispersion \citep{Hoagland1985, Adrover2019}, we will define a new effective dispersion coefficient for the long-term (many period) growth of the axial variance $D_{\mathrm{eff}}^{\infty}$ defined as follows:
\begin{equation}
D_{\mathrm{eff}}^{\infty}\equiv\lim_{t\rightarrow\infty}\frac{\sigma_x^2}{2t}
\end{equation}
The non-dimensional version of this quantity will be defined as $D_{\mathrm{eff}}^{\infty*}\equiv D_{\mathrm{eff}}^{\infty}/D$. We analyzed long-term dispersion behavior in periodic channels with three different shapes but with the same period and similar radius amplitude. Informed by our analysis in the previous section, however, the predicted variance growth becomes inaccurate when the ratio $\sigma_x/\lambda$ is greater than about 0.3. We thus computed the long-term growth of the axial variance from our current model as follows:
\begin{equation}
    D_{\mathrm{eff, curr}}^{\infty}=\frac{1}{\left.t\right|_{\sigma_x=0.3\lambda}}\int_{0}^{\left.t\right|_{\sigma_x=0.3\lambda}}\frac{\sigma_x^2(t)}{2t}\mathrm{d}t
\end{equation}

Figure \ref{fig:literature} summarizes the results of this study. The three plots in the top row show plots of the radius distribution $a(x)$ normalized by initial radius $a_0$ as a function of the normalized axial coordinate $x^*$. The three functions $a(x)$ are sinusoidal, triangular, and exponential-sine-wave functions of the following forms:
\begin{equation}
\begin{split}
a_1(x)&=1+\delta\sin(2\pi x/\lambda)\\
a_2(x)&=1+\frac{3\delta}{2\lambda}\mathrm{mod}(x,\lambda)-\frac{9\delta}{2\lambda}(\mathrm{mod}(x,\lambda)-\frac{2}{3}\lambda)H(\mathrm{mod}(x,\lambda)-\frac{2}{3}\lambda)\\
a_3(x)&=\frac{\delta}{e}\exp(\sin(2\pi x/\lambda)) + (1-\frac{\delta}{e}),
\end{split}
\end{equation}
where $H(x)$ is the Heaviside step function. The three plots in the second row of Figure \ref{fig:literature} show the effective, long-term dispersion coefficient $D_{\mathrm{eff}}^{\infty*}$ for each case as a function of $Pe_{a_0}$. $D_{\mathrm{eff}}^{\infty*}$ curves are shown for the current model (data point) and Brownian simulations (solid line). For these conditions, we see excellent agreement between the current model and the Brownian simulations for the long-term dispersion coefficient. The results capture an asymptote of $D_{\mathrm{eff}}^{\infty*}=1$ for vanishing $Pe_{a_0}$, as expected. $D_{\mathrm{eff}}^{\infty*}$ increases monotonically with increasing $Pe_{a_0}$ and asymptotes to a $Pe_{a_0}^2$ dependence for large $Pe_{a_0}$.

In the left-hand plot of the bottom row, we also show a comparison of our model results with the work of Adrover et al. \citep{Adrover2019}. Adrover analyzed the long-term dispersion of solutes in a sinusoidal channel and provided an approximate analytical formula for $D_{\mathrm{eff}}^{\infty*}$ as a function of $Pe_{a_0}$, and this prediction is plotted along with our model in the figure. Adrover's also found that $D_{\mathrm{eff}}^{\infty*}$ tends to unity for small $Pe_{a_0}$ and scales with the second power of $Pe_{a_0}$ for large $Pe_{a_0}$. We note the excellent agreement among the three predictions.

We conclude the current model can be readily adapted to a wide variety of periodic channel shapes. We further note the similarity among the three $D_{\mathrm{eff}}^{\infty*}$ versus $Pe_{a_0}$ curves for the three cases. This similarity leads us to hypothesize that the long-term solute dispersion in such periodic channels is largely driven by the spatial frequency and the amplitude of the radius oscillation (and $Pe_{a_0}$) and may be insensitive to the details of channel shapes.

\subsection{Arbitrarily-shaped channels}
\begin{figure}
\centerline{\includegraphics[width=1.0\textwidth]{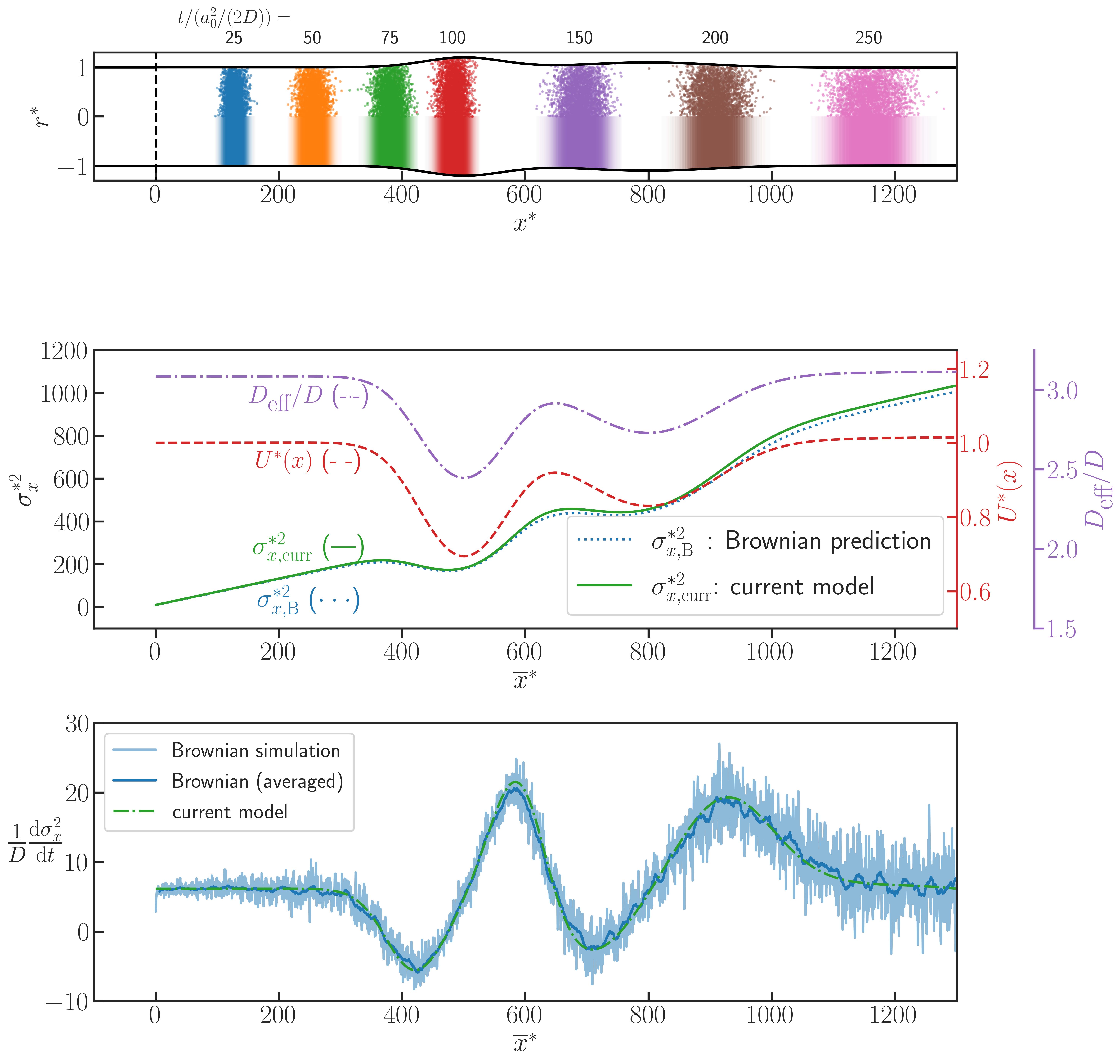}}
\caption{Taylor-Aris dispersion for an example arbitrarily shaped axisymmetric channel. The top plot shows results from a Brownian dynamic simulation (upper half) and axial solute distribution predicted with the current model (Equation \ref{eq:var}). The middle plot shows distribution of $\sigma_{x,\mathrm{B}}^{*2}$, $\sigma_{x,\mathrm{curr}}^{*2}$, $U^*(x)$, and $D_{\mathrm{eff}}/D$, each as a function of the non-dimensional axial location along the channel. $U^*(x)$ and $D_{\mathrm{eff}}/D$ are shown as a reference. Variance computed using Equation \ref{eq:var} shows excellent agreement with Brownian dynamics simulations. The bottom plot shows the same example used for demonstration and benchmarking of Equation \ref{eq:regime}. Plotted is a scaled rate of change of axial variance as a function of mean solute position $\overline{x}$ as computed using Equation \ref{eq:var} and analytical expression Equation \ref{eq:regime}. Both of these solutions are compared to calculations based on a Brownian dynamics simulation. The blue curve is the Brownian smoothed with a moving average with a windows size of $\Delta t^*=2.5$, while the light blue curve is the original Brownian simulation results. 
}\label{fig:arbitrary}
\end{figure}

We next demonstrate a novel application of our model to a particular but arbitrary axisymmetric channel shape. Similar to the top plot of Figure \ref{fig:periodic}, the top plot of Figure \ref{fig:arbitrary} shows a plot of the channel geometry in coordinates of $r^*$ versus $x^*$ in a highly exaggerated aspect ratio for clarity. We show results for $Pe_{a_0}$ of 10 where $a_0$ is the radius at $x=0$. The channel shows plots of solute zones at non-dimensional times ranging from 25 to 250. The top half of the solute zones are raw results from the Brownian dynamics simulations, while the bottom half are plots of the predicted axial distribution of area-averaged concentration from the current model. Again we see excellent agreement between the axial distributions of Brownian dynamics particle densities and the axial distributions from the model. Note how the model captures the sudden positive and negative growth of axial variance as the solute traverses through contractions and expansions, respectively. For example, note the sudden decreases in variance due to the expansion just upstream of $x^*=500$, and then the sudden increases in variance caused by the contraction near $x^*=700$.

The middle plot of Figure \ref{fig:arbitrary} shows the axial variance of the solute zone as a function of nondimensional solute average axial location, $\overline{x}^*$. Plotted are the axial variance from the Brownian dynamics simulation and the current model. Note the excellent agreement between these two models again showing how the current model captures very well the detailed development of the axial variance. For example, note the sudden decreases of axial variance just upstream of $\overline{x}^*=500$ and $\overline{x}^*=800$, and the sudden increases of axial variance just upstream of $\overline{x}^*=700$.

The middle plot of Figure \ref{fig:arbitrary} also shows the non-dimensional area-averaged flow velocity and the local normalized effective dispersion coefficient, $D_{\mathrm{eff}}/D$ (c.f. two ordinances on the RHS). Similar to the case in Figure \ref{fig:periodic}, both $U$ and $D_{\mathrm{eff}}^*$ show opposite trend to the shape of the channel. $D_{\mathrm{eff}}^*$ again increases as the channel constricts and decreases as the channel expands, as we expected. However, even in a region where axial variance is transiently decreasing (e.g. between $x^*=400$ to $500$), the effective dispersion coefficient remains positive and greater than unity. This again highlights the inability of effective dispersion coefficient to describe the variance evolution, here in an arbitrarily-shaped axisymmetric channel.

The bottom plot of Figure \ref{fig:arbitrary} shows the scaled rate of change of variance ($\frac{1}{D}\frac{\mathrm{d}\sigma_x^2}{\mathrm{d}t}$) as a function of solute average axial location $\overline{x}^*$. Plotted are the results computed from Brownian simulations and the current model (Eq.\ref{eq:var}). For the Brownian dynamics simulation, both raw data and the moving averaged results are shown. The axial variance growth rate varies according to the geometry of the channel (shown in the top plot of the same figure), decreasing when the channel expands and increasing when the channel contracts, as we expect. There is also excellent agreement among the three solutions, and the current model captured the negative variance growth rate near $\overline{x}^*=400$ and $\overline{x}^*=700$. This shows the capability of our model to predict the axial variance evolution, and also provides validation of using Equation \ref{eq:regime} to identify the regime of transient negative axial variance growth.

\subsection{Engineering channel shape for specific variance patterns}
We apply our analytical approach to design the channel shape to have a desired spatial evolution of axial variance. As a proof of concept, we design one channel that maintains an approximately constant axial variance (channel A) and a second channel that results in a sinusoidal variation of axial variance as the solute develops in the channel (channel B). 

\begin{figure}
\centerline{\includegraphics[width=1.0\textwidth]{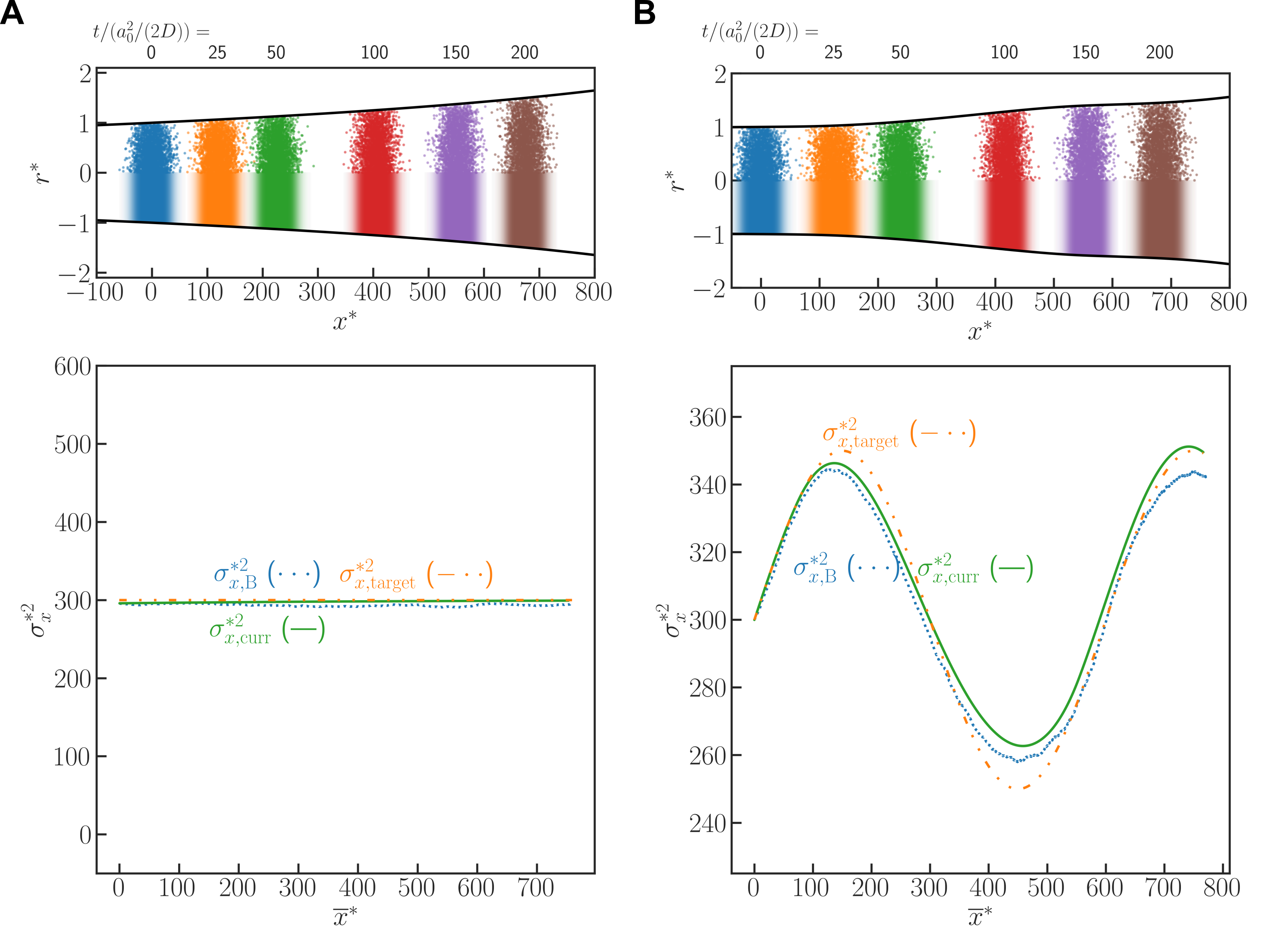}}
\caption{Engineering the axial variance evolution in Taylor-Aris dispersion. Using Equation \ref{eq:arb_var}, we designed two channels which (A) maintain an approximately constant axial variance and (B) result in a sinusoidal (axial) variation of axial variance as the solute develops in the channel. The top plot shows results from a Brownian dynamic simulation (upper half) and axial solute distribution predicted with the current model (Equation \ref{eq:var}). The bottom plot shows distribution of $\sigma_{x,\mathrm{B}}^{*2}$, $\sigma_{x,\mathrm{curr}}^{*2}$, and $\sigma_{x,\mathrm{target}}^{*2}$, each as a function of the non-dimensional axial location along the channel. Axial variance computed using Equation \ref{eq:var} and axial variance computed from Brownian dynamic simulation both show excellent agreement with the targeted variance evolution pattern.
}\label{fig:engineer}
\end{figure}

The two channel shapes are designed by solving Equation \ref{eq:arb_var}. The shape of channel A monotonically diverges, and the rate of diverging increases downstream. For channel B, there is an increased diverging rate between $x^*=300$ and $500$, coinciding with the region where it is necessary to reduce the variance according to the targeted sinusoidal pattern. Note that the channel shape in Figure \ref{fig:engineer}A has an analytical expression. By setting the $\frac{\mathrm{d}\sigma_x^2}{\mathrm{d}x}$ term on the RHS of Equation \ref{eq:arb_var} to 0, we can simplify Equation \ref{eq:arb_var} into the following ODE for $a(x)$:
\begin{equation}\label{eq:const_var}
\frac{\mathrm{d}a}{\mathrm{d}x}\approx \frac{D}{2\sigma_x^2U_0a_0^2}a^3+\frac{U_0a_0^2}{96\sigma_x^2D}a
\end{equation}
which has an analytical solution as follows:
\begin{equation}\label{eq:const_var_sol}
a^2(x)=\frac{\frac{U_0a_0^2}{96\sigma_x^2D}e^{\frac{U_0a_0^2}{48\sigma_x^2D}(c_1+x)}}{1-\frac{D}{2\sigma_x^2U_0a_0^2}e^{\frac{U_0a_0^2}{48\sigma_x^2D}(c_1+x)}}
\end{equation}
where $c_1$ is the constant of integration related to the initial variance. We know of no analytical solution for the equation for the case of channel B.

Figures \ref{fig:engineer}A and \ref{fig:engineer}B show solutions for dispersion of solute zone in the two engineered channels. The channel in Figure \ref{fig:engineer}A (abbreviated as "channel A") is designed to maintain an approximately constant axial variance of 300 ($\sigma_x^{2*}(\overline{x}^*)=300$), while the channel in Figure \ref{fig:engineer}B (abbreviated as "channel B") is designed to yield a sinusoidal axial variation of axial variance ($\sigma_x^{2*}(\overline{x}^*)=300+50\sin(2\pi\overline{x}^*/600)$). Similar to the top plot of Figure \ref{fig:divconv}, the top two plots of Figure \ref{fig:engineer} shows the two engineered channel geometry in coordinates of $r^*$ versus $x^*$. We show results for $Pe_{a_0}$ of 10 where $a_0$ is taken as the radius at $x=0$. The top half of the solute concentration in the channel are raw data results from Brownian dynamics simulations, while the bottom half of the channels show the area-averaged axial distribution of solute concentration predicted from the current model. 

The bottom two plots in Figure \ref{fig:engineer} are plots of the axial variance of the area-averaged axial distribution of solute concentration from the current model ($\sigma_{x, \mathrm{curr}}^{*2}$), Brownian dynamics ($\sigma_{x, \mathrm{B}}^{*2}$), and the targeted axial variance spatial evolution pattern ($\sigma_{x, \mathrm{target}}^{*2}$). These quantities are plotted as a function of the axial mean position of the solute zone, $\overline{x}^*$. For channel A, there is a near-perfect agreement among the current model, Brownian dynamics simulation, and targeted pattern. All three lines stay flat at a variance of 300 as we expected. For channel B, there is a good agreement among the three quantities. Although the amplitude of the sinusoidal variation is slightly smaller compared to the targeted pattern, the Brownian dynamics simulation results show the desired sinusoidal spatial evolution pattern. To our knowledge, our study is the first to demonstrate the design of a channel shape that yields a desired spatial evolution pattern of variance. Without the analytical approach described in Section 2.2-2.4, this process would require repetitive simulation efforts (as with shape optimization) to compute channel shapes to obtain the desired pattern. We hypothesize that the proposed engineered channel shape can be produced by additive manufacturing methods.

Also, the most common application of microfluidics is the chemical and biochemical analyses of species \citep{Wu2016}. The most common method of detection of analytes is an optical detection technique such as fluorescence, colorimetric, or UV adsorption \citep{Wu2011}. These optical techniques perform line-of-sight averaging of solutes in long-thin channels. Hence, the detection of species is typically proportional to the area average of some analyte solute zone - and this is the main characteristic of interest of the current work. The ability to tailor a channel shape so as to preserve the area-averaged concentration therefore offers the opportunity to tailor microchannel shapes that can transport species via pressure-driven flow while also preserving and/or tailoring the depth- or area-averaged signal strength.

\section{Summary and conclusions}
We demonstrated a Taylor-Aris dispersion analysis for axisymmetric channels with slowly-varying, arbitrary radius distributions, with assumptions and smallness parameters summarized in Table \ref{table:assumptions_summary}. We first derived a PDE for the development of the area-averaged concentration including an explicit local dispersion coefficient. We then derived equations for the dynamics of the axial mean (first moment) and variance (second moment) of the solute distribution. We then proposed a heuristic for relations describing the moments of local geometry experienced by the solute. Our analysis allowed us to simplify the solution for the complex dynamics of solute zones to two coupled ODEs for the mean and variance. These ODEs provide a description of solute position and variance from the channel geometry, given the assumptions of lubrication flow. To our knowledge, this is the first time a full prediction of the time evolution of axial variance is possible using only two ODEs (including for short time scales) for this type of problem. Our method can also be applied to a long time scale, as long as the axial variance remains small compared to the characteristics wavelength of channel variations.

We further derived an analytical expression which delineates the regimes of transient positive and negative axial variance growth. This expression quantifies the solution of transient zero growth axial variance as a function of the local Peclet number, local slope in the channel, and the ratio between axial variance and the square of the local radius. This analysis demonstrates clearly the conditions required for channel expansion to yield decreases in solute axial variance. We also developed further simplifications of our model which yields a single, first-order nonlinear ODE describing the relation between the axial radius distribution and axial variance (spatial) distribution. This relation is very useful in the design of channel shapes which yield specific (desired) dynamics for solute axial variance.

We applied our model to several interesting test cases and benchmarked its performance relative to Brownian dynamics simulations. First, we demonstrated our model yields accurate predictions (relative to Brownian dynamics) of area-averaged solute dynamics in diverging and converging (conical) channels. Second, we applied our model to predict solute dynamics for various periodic channels, and again demonstrated excellent agreement with Brownian dynamics simulations. For the latter, we considered an initial condition and regime where channel expansions result in substantial decreases in axial variance (i.e. transient negative solute axial variance growth). We increased the Peclet number of the channel from 10 to 1000 and studied how the performance of our model fails as the key assumptions are violated. We further analyzed the long-term (many period) developments of solute in periodic channels by defining a long-term effective dispersion coefficient. We analyzed three separate periodic channel shapes and showed this long-term behavior is affected mostly by channel (axial) period and the magnitude of the fluctuation of the radius function. For the case of sinusoidal channels, our model demonstrated excellent comparison with a previously published model.

The third example application of our model was for an arbitrarily shaped channel. We here selected some complex channel shape which demonstrated strong advective dispersion effects. Again, our model showed excellent comparison with Brownian dynamics simulations, including the capture of the positive and transient negative growth in axial variance when the solute zone experiences constriction or expansion in the channel.

Lastly, we demonstrated the power of our analytical approach by designing two channels that can control the spatial evolution of the solute so as to produce a desired spatial distribution of the solute axial variance. For the latter work, we first designed a channel that can maintain an approximately constant solute axial variance. We then demonstrated a channel geometry which produces a sinusoidal axial distribution of axial variance as the solute develops in the channel.

Overall, our analysis provides a fairly accurate (according to Brownian dynamics simulation), fast, and easy-to-use model for solute dynamics in axisymmetric channels of arbitrary variance.

\section{Acknowledgements}
R.C. gratefully acknowledges support from the Stanford University Bio-X SIGF Fellows Program and from the Ministry of Education in Taiwan.

\section{Declaration of Interests}
The authors report no conflict of interest.

\bibliographystyle{jfm}
\bibliography{references}
\end{document}


\maketitle
\tableofcontents
\newpage

\section{Intermediate steps in deriving Eq. 2.19}
In Eq. 2.18 of the main manuscript we derived an expression of $c^{\prime}(x,r,t)$ in terms of $\frac{\partial\langle c\rangle}{\partial x}$. We here present some details of the derivation. The followings are expressions for each of the terms in Eq. 2.11:
\begin{equation}
\begin{split}
\frac{\mathrm{d}U}{\mathrm{d}x}=U_0a_0^2\frac{\mathrm{d}}{\mathrm{d}x}\left(\frac{1}{a^2}\right)=-2U_0a_0^2\frac{\beta}{a^3}=\frac{-2Ua^2\beta}{a^3}=\frac{-2U\beta}{a}
\end{split}
\end{equation}
\begin{equation}
\begin{split}
\left.c^{\prime}\right|_{r=a} =\frac{1}{24}\frac{Ua^2}{D}\frac{\partial\langle c\rangle}{\partial x}+\frac{1}{4}\beta a\frac{\partial\langle c\rangle}{\partial x}
\end{split}
\end{equation}
\begin{equation}
\begin{split}
\frac{\partial c^{\prime}}{\partial x} &= \left\{\frac{U\beta}{2aD}\left(-r^2+\frac{r^4}{a^2}\right) + \left(-\frac{\beta^2r^2}{2a^2}+\frac{\gamma r^2}{2a} -\frac{1}{4}\gamma a - \frac{1}{4}\beta^2\right) \right\}\frac{\partial\langle c\rangle}{\partial x}\\
&+ \left\{\frac{U}{D}\left(\frac{r^2}{4}-\frac{r^4}{8a^2}-\frac{a^2}{12}\right)+\left(\frac{\beta r^2}{2a}-\frac{\beta a}{4}\right)\right\}\frac{\partial^2\langle c\rangle}{\partial x^2}
\end{split}
\end{equation}
\begin{equation}
\begin{split}
\int_{0}^{a}c^{\prime}\mathrm{d}r&=\int_{0}^{a}\left[\frac{\partial \langle c\rangle}{\partial x}\frac{U(x)}{D}\left(\frac{r^2}{4} - \frac{r^4}{8a^2} -\frac{a^2}{12}\right)+\left(\frac{\beta r^2}{2a} - \frac{1}{4}\beta a\right)\frac{\partial \langle c\rangle}{\partial x} \right]\mathrm{d}r\\&=-\frac{1}{40}\frac{Ua^3}{D}\frac{\partial\langle c\rangle}{\partial x}-\frac{1}{12}\beta a^2 \frac{\partial\langle c\rangle}{\partial x}
\end{split}
\end{equation}
\begin{equation}
\begin{split}
\left.\frac{\partial c^{\prime}}{\partial r}\right|_{r=a} = \frac{U}{D}\left(\frac{a}{2}-\frac{a}{2}\right)\frac{\partial\langle c\rangle}{\partial x}+\beta\frac{\partial\langle c\rangle}{\partial x}=\beta\frac{\partial\langle c\rangle}{\partial x}
\end{split}
\end{equation}
\begin{equation}
\begin{split}
\frac{\partial}{\partial x}\left(a\left.c^{\prime}\right|_{r=a}\right)&=\frac{\partial}{\partial x}\left(\left(\frac{1}{24}\frac{U_0a_0^2}{D}a + \frac{1}{4}\beta a^2 \right)\frac{\partial\langle c\rangle}{\partial x}\right)\\
&=\left(\frac{U_0a_0^2\beta}{24D} + \frac{\gamma a^2}{4} + \frac{\beta^2 a}{2}\right)\frac{\partial\langle c\rangle}{\partial x}+\left(\frac{Ua^3}{24D}+\frac{\beta a^2}{4}\right)\frac{\partial^2\langle c\rangle}{\partial x^2}
\end{split}
\end{equation}
\begin{equation}
\begin{split}
\left\langle u_r^{\prime}\frac{\partial c^{\prime}}{\partial r}\right\rangle &=\frac{2}{a^2}\int_{0}^{a}r\left(2\beta U\left(\frac{r}{a}-\frac{r^3}{a^3}-\frac{4}{15}\right)\left[\frac{U}{D}\left(\frac{r}{2}-\frac{r^3}{2a^2}\right)+\frac{\beta r}{a}\right]\frac{\partial\langle c\rangle}{\partial x} \right)\mathrm{d}r\\&=\frac{4\beta U}{a^2}\frac{\partial\langle c\rangle}{\partial x}\left(\frac{11}{3600}\frac{Ua^3}{D}-\frac{\beta a^2}{180}\right)
\end{split}
\end{equation}
\begin{equation}
\begin{split}
\left\langle u_x^{\prime}\frac{\partial c^{\prime}}{\partial x}\right\rangle &=\frac{2}{a^2}\int_{0}^{a}U\left(1-\frac{2r^2}{a^2}\right)\\&[\left\{\frac{U\beta}{2aD}\left(-r^2+\frac{r^4}{a^2}\right)+\left(\frac{-\beta^2r^2}{2a^2}+\frac{\gamma r^2}{2a}-\frac{\gamma a}{4}-\frac{\beta^2}{4}\right) \right\}\frac{\partial\langle c\rangle}{\partial x} \\&+ \left\{ \frac{U}{D}\left(\frac{r^2}{4}-\frac{r^4}{8a^2}-\frac{a^2}{12}\right) + \left(\frac{\beta r^2}{2a}-\frac{\beta a}{4}\right)\right\}\frac{\partial^2\langle c\rangle}{\partial x^2} ] r\mathrm{d}r\\&=\frac{U}{12}(\beta^2-a\gamma)\frac{\partial\langle c\rangle}{\partial x}-\frac{a^2U^2}{48D}\frac{\partial^2\langle c\rangle}{\partial x^2}-\frac{Ua\beta}{12}\frac{\partial^2\langle c\rangle}{\partial x^2}
\end{split}
\end{equation}
Substituting each term into Eq. 2.11 and we arrive at Eq. 2.19 of the main manuscript.
\newpage
\section{Derivation of the dominant error term in  Eq. 2.31}
As stated in the main manuscript, the ODE for the dynamics of $\sigma_x^2$ can be written as $\frac{\mathrm{d}\sigma^2_x}{\mathrm{d}t}=2Ua^2\left[\overline{\frac{x}{a^2}}-\overline{x}\overline{\frac{1}{a^2}}\right]+\frac{U^2a^4}{24D}\overline{\frac{1}{a^2}}+2D+4D\overline{\frac{\beta}{a}(x-\overline{x})}$. The dominant error term listed in Eq. 2.31 comes from the first term $2Ua^2\left[\overline{\frac{x}{a^2}}-\overline{x}\overline{\frac{1}{a^2}}\right]$. Taylor expansion of this term according to Eqs. 2.26 and 2.29 yields:
\begin{equation}
\begin{split}
\left[\overline{\frac{x}{a^2}}-\overline{x}\overline{\frac{1}{a^2}}\right] &= \overline{\overline{x}\widetilde{a^{-2}}} + \overline{(x-\overline{x})(\overline{x}\widetilde{(a^{-2})^{\prime}} + \widetilde{a^{-2}})} + \overline{(x-\overline{x})^2\left(\frac{1}{2}\overline{x}\widetilde{(a^{-2})^{\prime\prime}}+\widetilde{(a^{-2})^{\prime}}\right)}\\
&+\overline{\frac{1}{6}(x-\overline{x})^3\left(\overline{x}\widetilde{(a^{-2})^{\prime\prime\prime}}+3\widetilde{(a^{-2})^{\prime\prime}}\right)} - \overline{x}\widetilde{a^{-2}} - \overline{(x-\overline{x})\widetilde{(a^{-2})^{\prime}}} \\
&- \overline{\frac{1}{2}(x-\overline{x})^2\widetilde{(a^{-2})^{\prime\prime}}}\overline{x} - \overline{\frac{1}{6}(x-\overline{x})^3\widetilde{(a^{-2})^{\prime\prime\prime}}}\overline{x} + \mathcal{O}(\sigma_x^4\mathrm{Kurt}_x\widetilde{(a^{-2})^{\prime\prime\prime}})\\
&= \sigma_x^2\widetilde{(a^{-2})^{\prime}}+\frac{1}{2}\sigma_x^3\mathrm{Skew}_x\widetilde{(a^{-2})^{\prime\prime}}+ \mathcal{O}(\sigma_x^4\mathrm{Kurt}_x\widetilde{(a^{-2})^{\prime\prime\prime}})\\
&= -2\sigma_x^2\widetilde{a^{-3}\beta} + \sigma_x^3\mathrm{Skew}_x\widetilde{\frac{3\beta^2-a\gamma}{a^4}}+ \mathcal{O}(\sigma_x^4\mathrm{Kurt}_x\widetilde{(a^{-2})^{\prime\prime\prime}}).
\end{split}
\end{equation}
Here $(\cdot)^{\prime}=\frac{\mathrm{d}(\cdot)}{\mathrm{d}x}$; $\mathrm{Skew}_x$ is the axial skewness of the solute zone, defined as $\mathrm{Skew}_x=\overline{(x-\overline{x})^3}/\sigma_x^3$; $\mathrm{Kurt}_x$ is the axial kurtosis of the solute zone, defined as $\mathrm{Kurt}_x=\overline{(x-\overline{x})^4}/\sigma_x^4$. The second term gives rise to the error listed in Eq. 2.31 of the main manuscript.

\newpage
\section{Summary of simulation parameters}
This section lists the parameters used for the Brownian dynamics simulations in the main manuscript.

\begin{table}
\small
\caption{Parameters for Brownian dynamics simulation.} 
\centering 
\begin{tabular}{ c } 
\\[0cm]
\makecell[l]{\textbf{- Shared parameters:}
\\$U_0 = 1$, $a_0 = 1$, $D = U_0a_0/Pe_{a_{0}}$.
\\Time step $\mathrm{d}t = Pe_{a_{0}}/40$. Number of timesteps = $40Nt_0+1$.
\\Number of points = 5000.
\\All simulations repeated with random seed 0 - 4.}\\
\\[0cm]
\makecell[l]{\textbf{- Figure 3:}
\\diverging: $a(x) = a_0 + \beta x$, $a_0 = 1$, $\beta = 10^{-3}$.
\\$Pe_{a_{0}} = 10$, $Nt_0$ = 500, $\sigma_{x0}^2 = 10$
\\converging: $a(x) = a_0 + \beta x$, $a_0 = 1$, $\beta = -10^{-3}$.
\\$Pe_{a_{0}} = 10$, $Nt_0$ = 26, $\sigma_{x0}^2 = 10$
} \\ 
\\[0cm]
\makecell[l]{\textbf{- Figure 4-6, S1:}
\\$a(x)=1+0.2\sin(x/400)$, $Nt_0$ = 500, $\sigma_{x0}^2 = 10$
\\Figure 4: $Pe_{a_{0}} = 10$
\\Figure 5: $Pe_{a_{0}} = 100$
\\Figure 6: $Pe_{a_{0}} = 1000$
}\\
\\[0cm]
\makecell[l]{\textbf{- Figure 7:}
\\$\lambda = 200$, $\delta = 0.05$, $Nt_0$ = 500, $\sigma_{x0}^2 = 10$
\\$Pe_{a_{0}} = 0.1, 1, 10, 100, 1000$
\\Left: $1+\delta\sin(2\pi x/\lambda)$
\\Middle: $1+\frac{3\delta}{2\lambda}\mathrm{mod}(x,\lambda) - \frac{9\delta}{2\lambda}(\mathrm{mod}(x,\lambda)-\frac{2}{3}\lambda)H(\mathrm{mod}(x,\lambda)-\frac{2}{3}\lambda)$
\\Right: $\frac{\delta}{e}\exp(\sin(2\pi x/\lambda)) + (1-\frac{\delta}{e})$
}\\
\\[0cm] 
\makecell[l]{\textbf{- Figure 8:}
\\$a(x) = 1+0.2e^{-(x-500)^2/8000} + 0.1e^{-(x-800)^2/20000} + 0.2 e^{-(x-2000)^2/100000} - 0.1e^{-(x-3500)^2/2000000}$
\\$Pe_{a_{0}} = 10$, $Nt_0$ = 500, $\sigma_{x0}^2 = 10$
}\\
\\[0cm]
\makecell[l]{\textbf{- Figure 9:}
\\constant axial variance:
\\$Pe_{a_{0}} = 10$, $Nt_0$ = 120,$\sigma_{x0}^2 = 300$
\\$A = \frac{D}{2\sigma_{x0}^2U_0a_0^2}$, $B = \frac{U_0a_0^2}{96\sigma_{x0}^2D}$, $c_1 = \frac{48\sigma_{x0}^2D}{U_0a_0^2}\ln\left(\frac{a_0^2}{(B+\frac{D}{2\sigma_{x0}^2U_0})}\right)$
\\$x(a) = \frac{1}{2B}\ln\left(\frac{a^2}{Aa^2+B} \right)-c_1$
\\sinusoidal axial variance:
\\$Pe_{a_{0}} = 10$, $Nt_0$ = 120,$\sigma_{x0}^2 = 300$.
\\$\delta=50$, $\lambda = 600$, $\sigma_x^{2}(\overline{x})=300+\delta\sin(2\pi\overline{x}/\lambda)$.
\\Solve Equation 2.35.
}\\
\\[0cm]
\end{tabular}
\label{table:numeircs_summary} 
\end{table}

\newpage
\section{Supplementary figures}
Figure S1 shows plots of the Brownian dynamics simulations (top) and the concentration distribution weighted by local cross-sectional area (bottom) for the channel with a sinusoidal radius distribution (the same channel as in Figure 4-6), with initial Peclet number $Pe_{a_{0}}$ varying from 10, 100 to 1000 and the mean axial location $\overline{x}$ the same. As expected, the solute zone subjected to a higher Peclet number spreads faster, and the axial concentration distribution is more skewed.

\begin{figure}
\centerline{\includegraphics[width=0.9\textwidth]{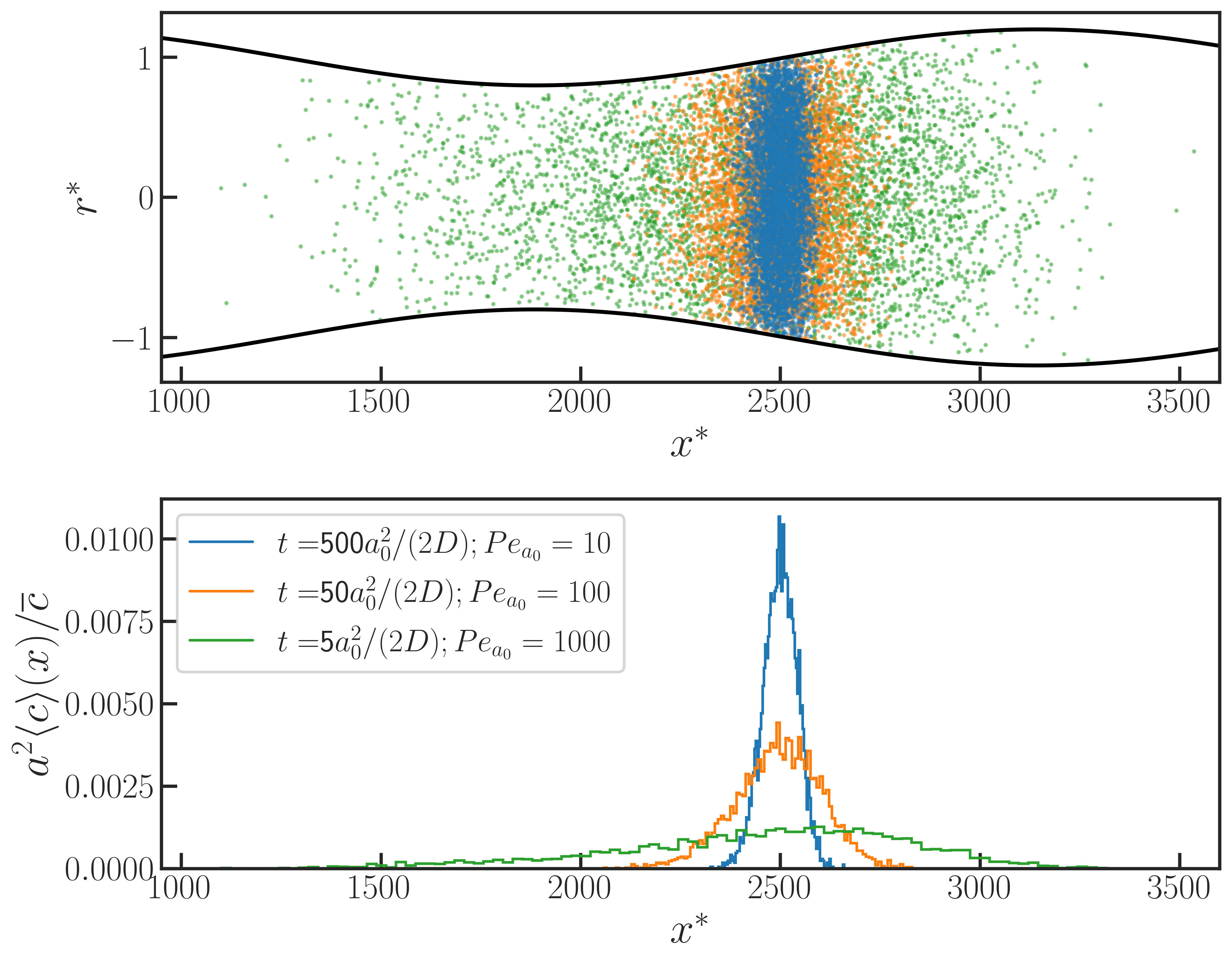}}
\caption{Taylor-Aris dispersion in a channel with a sinusoidal (periodic) radius distribution, $a(x)$. The channel is the same as that of Figure 4-6, with initial Peclet numbers $Pe_{a_{0}}$ of 10 (blue, $t=500a_0^2/(2D)$), 100 (orange, $t=50a_0^2/(2D)$) to 1000 (green, $t=5a_0^2/(2D)$). The top plot shows results from a Brownian dynamic simulation. The bottom plot shows the concentration distribution normalized by local cross-sectional area. For a fixed distance of travel, the solute zones subjected to greater Peclet numbers disperse faster and the solute distribution becomes more skewed.
}\label{fig:Pe_skewness}
\end{figure}

\newpage
\section{Comparison of effective dispersion coefficient in periodic channels with solutions by Adrover et al.}
Adrover et al. comprehensively analyzed effective dispersion coefficient in a wide range of wavelengths for a periodic channel with a specific radius $R(x)=R_0(1+\delta\sin(2\pi x/(\lambda R_0)))$ \citep{Adrover2019}. For our current model, the Taylor dispersion analysis requires that the flow field in the channel be well approximated by lubrication theory. This constraint in turn requires the ratio ($\lambda$) of the channel wavelength to channel radius be significantly greater than unity. In this limit, Adrover et al. provided the following asymptotic formula for the effective dispersion coefficient in the axisymmetric sine-wave channel as
\begin{equation}\label{eq:Deff_lit}
\begin{split}
    D_{\mathrm{eff}}^{\infty*}=D_0^* + \epsilon^2 D_1^{*}+\epsilon^4 D_2^{*}+...\\
    \lim_{Pe_{\lambda}\rightarrow\infty}D_0^*=\frac{16+120\delta^2+90\delta^4+5\delta^6}{16(1+\delta^2/2)}\\
    D_1^{*}=\frac{Pe_{\lambda}^2}{48}\frac{(1+3\delta^2+3\delta^4/8)}{(1+\delta^2/2)^3}\\
    D_2^{*}=Pe_{\lambda}^2\frac{43\pi^2\delta^2(1+3\delta^2/2+\delta^4/8)}{480(1+\delta^2/2)^3}
\end{split}
\end{equation}
Here, $\epsilon^2=\lambda^{-2}$ and $Pe_{\lambda}=\lambda\frac{UR_0}{D}$ is the Peclet number defined based on wavelength. 
\newpage

\bibliographystyle{jfm}
\bibliography{references}